\documentclass[letterpaper, 10 pt, journal, twoside]{IEEEtran}
\pdfoutput=1
\usepackage{hyperref}
\hypersetup{hidelinks,pdfnewwindow,breaklinks} 
\usepackage{cite}
\usepackage{amsmath,amssymb,amsfonts}
\usepackage{algorithmic}
\usepackage{graphicx}
\usepackage{textcomp}
\usepackage{bm}
\usepackage{listings}

\usepackage[caption=false,font=normalsize,labelfont=sf,textfont=sf]{subfig}
\usepackage{mathtools}
\usepackage{here}
\usepackage{color}
\usepackage{import}
\usepackage{mathrsfs}
\definecolor{caltechblue}{rgb}{0,0.2314,0.2980}
\definecolor{caltechgreen}{rgb}{0,0.3451,0.3137}
\definecolor{caltechorange}{rgb}{1,0.4235,0.0406}

\usepackage{boxedminipage}
\makeatletter
\def\ps@IEEEtitlepagestyle{
\def\@oddhead{\hbox{}\@IEEEheaderstyle\leftmark\hfil\thepage}\relax
\def\@evenhead{\@IEEEheaderstyle\thepage\hfil\leftmark\hbox{}}\relax
  \def\@oddfoot{\mycopyrightnotice}
  \def\@evenfoot{}
}
\def\mycopyrightnotice{
  {\footnotesize
  \begin{boxedminipage}{\textwidth}
  \centering
  © 2020 IEEE. Personal use of this material is permitted. Permission from IEEE must be obtained for all other uses, in any current or future media, including reprinting/republishing this material for advertising or promotional purposes, creating new collective works, for resale or redistribution to servers or lists, or reuse of any copyrighted component of this work in other works. Digital Object Identifier (DOI): {\color{caltechgreen}\underline{\href{https://ieeexplore.ieee.org/document/9261103}{10.1109/TAC.2020.3038402}}}
  \end{boxedminipage}
  }
}

\newtheorem{theorem}{Theorem}

\newtheorem{proposition}{Proposition}
 
\newtheorem{remark}{Remark}

\newtheorem{assumption}{Assumption}

\newtheorem{lemma}{Lemma}

\newtheorem{corollary}{Corollary}

\newcommand{\ie}{i.e.}

\newcommand{\st}{s.t.}

\DeclareMathOperator{\image}{Im}
\DeclareMathOperator{\kernel}{Ker}
\DeclareMathOperator{\trace}{Tr}
\DeclareMathOperator{\sym}{sym}

\def\BibTeX{{\rm B\kern-.05em{\sc i\kern-.025em b}\kern-.08em
    T\kern-.1667em\lower.7ex\hbox{E}\kern-.125emX}}
\title{Robust Controller Design for Stochastic Nonlinear Systems via Convex Optimization}
\author{Hiroyasu Tsukamoto, \IEEEmembership{Graduate Student Member, IEEE}, and Soon-Jo Chung, \IEEEmembership{Senior Member, IEEE}
\thanks{The authors are with the Graduate Aerospace Laboratories (GALCIT), California Institute of Technology, 1200 E California Blvd, Pasadena, CA, USA. E-mail: {{\tt\small\{htsukamoto, sjchung\}@caltech.edu}}, Code: {\color{caltechgreen}\underline{\url{https://github.com/astrohiro/cvstem}}}.}%
}

\begin{document}

\maketitle

\begin{abstract}
This paper presents \underline{C}on\underline{V}ex optimization-based \underline{S}tochastic steady-state \underline{T}racking \underline{E}rror \underline{M}inimization (CV-STEM), a new state feedback control framework for a class of It\^{o} stochastic nonlinear systems and Lagrangian systems. Its innovation lies in computing the control input by an optimal contraction metric, which greedily minimizes an upper bound of the steady-state mean squared tracking error of the system trajectories. Although the problem of minimizing the bound is non-convex, its equivalent convex formulation is proposed utilizing state-dependent coefficient parameterizations of the nonlinear system equation. It is shown using stochastic incremental contraction analysis that the CV-STEM provides a sufficient guarantee for exponential boundedness of the error for all time with $\bf{\mathcal{L}_2}$-robustness properties. For the sake of its sampling-based implementation, we present discrete-time stochastic contraction analysis with respect to a state- and time-dependent metric along with its explicit connection to continuous-time cases. We validate the superiority of the CV-STEM to PID, $\mathcal{H}_\infty$, and baseline nonlinear controllers for spacecraft attitude control and synchronization problems.
\end{abstract}
\section{Introduction}
\IEEEPARstart{S}{table} and optimal feedback control of It\^{o} stochastic nonlinear systems~\cite{nla.cat-vn712853} is an important, yet challenging problem in designing autonomous robotic explorers operating with sensor noise and external disturbances. Since the probability density function of stochastic processes governed by It\^{o} stochastic differential equations exhibits non-Gaussian behavior characterized by the Fokker-Plank equation~\cite{nla.cat-vn712853,PALLESCHI1990378}, feedback control schemes developed for deterministic nonlinear systems could fail to meet control performance specifications in the presence of stochastic disturbances.
\subsection{Contributions}
\begin{figure}
    \centering
    \includegraphics[width=85mm]{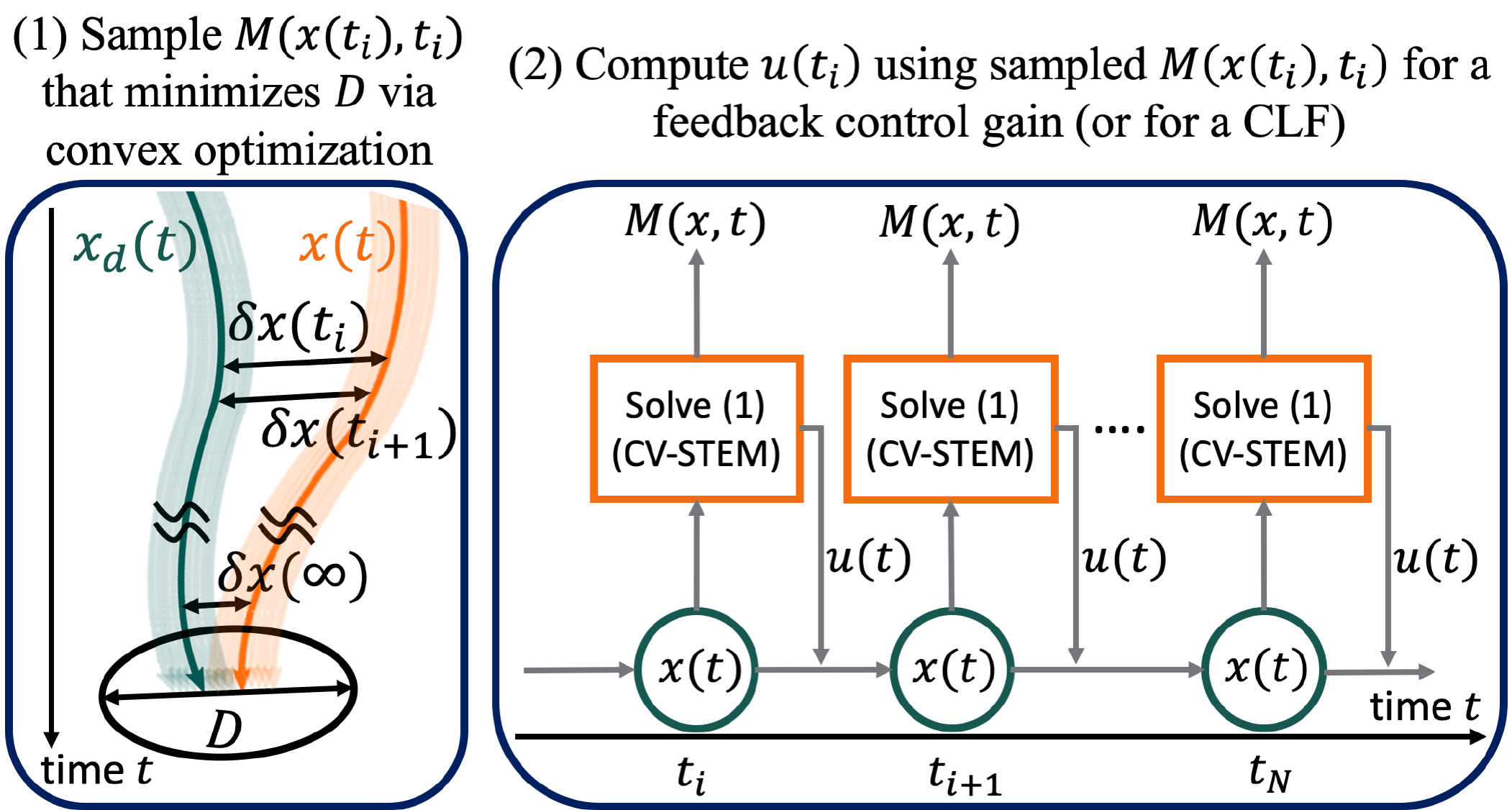}
    \caption{Illustration of the CV-STEM control: $M(x,t)$ denotes the optimal contraction metric for the differential Lyapunov function $\delta x^{\top} M(x,t)\delta x$; $x(t)$ and $x_d(t)$ are controlled and desired system trajectories; $u(t)$ is the control input computed by $M(x,t)$ (see Sec.~\ref{controller_design} for details).}
    \label{cvstemdrawing}
\end{figure}

The main purpose of this paper is to propose \underline{C}on\underline{V}ex optimization-based \underline{S}tochastic steady-state \underline{T}racking \underline{E}rror \underline{M}inimization (CV-STEM), a new framework to design an optimal contraction metric for feedback control of It\^{o} stochastic nonlinear systems and stochastic Lagrangian systems as in Fig.~\ref{cvstemdrawing}.
Contrary to Lyapunov theory, which gives a sufficient condition for exponential convergence, the existence of a contraction metric leads to a necessary and sufficient characterization of exponential incremental stability of nonlinear system trajectories~\cite{contraction,989067}.
We explore this approach further to obtain an optimal contraction metric for controlling It\^{o} stochastic nonlinear systems. This paper builds upon our prior work~\cite{mypaper}, but provides more rigorous proofs and explanations on how we convexify the problem of minimizing $D$ in Fig.~\ref{cvstemdrawing} in a mean squared sense. We also investigate its stochastic incremental stability properties and the impact of sampling-based implementation on its control performance both in detail, thereby introducing several additional theorems and simulation results. The construction and contributions of our CV-STEM method are summarized as follows.

1) The CV-STEM design is based on a convex combination of multiple State-Dependent Coefficient (SDC) forms of a nonlinear system equation (\ie{} $f(x,t)$ written as $A(x,t)x$~\cite{sddre,Banks2007,survey_SDRE}, where $A(x,t)$ is not necessarily unique). The main advantage of our control synthesis algorithm lies in solving an optimization problem, the objective of which is to find an optimal contraction metric that greedily minimizes an upper bound of the steady-state mean squared tracking error of It\^{o} stochastic nonlinear system trajectories, constructing an optimal feedback control gain and Control Lyapunov Function (CLF)~\cite{doi:10.1137/0321028,SONTAG1989117,Khalil:1173048} (see Fig.~\ref{cvstemdrawing}). Although the problem of minimizing the bound is originally non-convex, we reformulate it as a convex optimization problem with the State-Dependent Riccati Inequality (SDRI) constraint expressed as an LMI~\cite{lmi}, which can be solved by various computationally-efficient numerical methods~\cite{citeulike:163662,Ben-Tal:2001:LMC:502969,5447065,lmi}. We also propose one way to utilize non-unique choices of SDC forms for verifying the controllability of the system. This result is a significant improvement over the observer design~\cite{observer}, whose optimization-cost function uses a linear combination of observer parameters without accounting for the contraction constraint, which we express as an LMI~\cite{lmi} in this paper. This approach is further extended to the control of stochastic Lagrangian systems with a nominal exponentially stabilizing controller, and its superiority to the prior work~\cite{alt_controller,5068839}, PID, and $\mathcal{H}_{\infty}$ control~\cite{doi:10.1137/S0363012903423727,256331,159566} is shown using results of numerical simulations on spacecraft attitude control and synchronization.

2) It is proven using stochastic incremental contraction analysis that any solution trajectory under the CV-STEM feedback control exponentially converges to the desired trajectory in a mean squared sense with a non-vanishing error term (which will be minimized as explained above). It is also shown that the controller is robust against external deterministic disturbances which often appear in parametric uncertain systems, and that the tracking error has a finite $\mathcal{L}_2$ gain with respect to the noise and disturbances acting on the system. We note that the mean-square bound does not imply the asymptotic almost-sure bounds although finite time bounds could be obtained~\cite{nla.cat-vn712853,stochastic_contraction}, as the CV-STEM-based Lyapunov function is not a supermartingale due to the non-vanishing steady-state error term.

3) Discrete-time stochastic incremental contraction analysis with respect to a state- and time-dependent metric is derived for studying the effect of sampling-based implementation of the CV-STEM on its control performance. It is proven that stochastic incremental stability of discrete-time systems reduces to that of continuous-time systems if the time interval is sufficiently small. It is shown in the numerical simulations that the CV-STEM sampling period $\Delta t$ can be relaxed to $\Delta t \leq 25$~(s) for spacecraft attitude control and $\Delta t \leq 350$~(s) for spacecraft tracking and synchronization control without impairing its performance.

4) Some extensions of the CV-STEM are derived to explicitly incorporate input constraints and to avoid solving the convex optimization problem at every time instant.
\subsection{Related Work}
CLFs~\cite{doi:10.1137/0321028,SONTAG1989117,Khalil:1173048} as well as feedback linearization~\cite{Isidori:1995:NCS:545735,Slotine:1228283,Khalil:1173048} are among the most widely used tools for controlling nonlinear systems perturbed by deterministic disturbances. Since there is no general analytical scheme for finding a CLF, several techniques are proposed to find them utilizing some special structure of the systems in question~\cite{532343,Primbs99nonlinearoptimal,1425952,AYLWARD20082163,5443730}. The state-dependent Riccati equation method~\cite{sddre,Banks2007,survey_SDRE} can also be viewed as one of these techniques and is applicable to systems that are written in SDC linear structure. Building on these ideas for deterministic systems, a stochastic counterpart of the Lyapunov methods is proposed in~\cite{stochastic_lyapunov} to design CLF-based state and output feedback control of stochastic nonlinear systems~\cite{unknown_covariance,doi:10.1002/rnc.1255}. For a class of strict-feedback and output-feedback stochastic nonlinear systems, there exists a more systematic way of asymptotic stabilization in probability using a backstepping-based controller~\cite{DENG1997143,Deng1999}. However, one drawback of these approaches is that they are primarily directed toward stability with some implicit inverse optimality guarantees. 

Some theoretical methodologies have been developed to explicitly incorporate optimality into their feedback control formulation. These include $\mathcal{H}_{\infty}$ control~\cite{Basar1995,256331,159566}, which attempts to minimize the $\mathcal{H}_{\infty}$ norm for the sake of optimal disturbance attenuation. Although it is originally devised for linear systems~\cite{29425,85062,151101,599969,doi:10.1137/S0363012996301336,1259458}, its nonlinear analogues are obtained in~\cite{256331,159566} and then expanded to stochastic nonlinear systems~\cite{doi:10.1137/S0363012903423727} unifying the results on the $\mathcal{L}_2$ gain analysis based on the Hamilton-Jacobi equations and inequalities~\cite{Khalil:1173048}. Although we could design feedback control schemes optimally for specific types of systems such as Hamiltonian systems with stochastic disturbances~\cite{Zhu2006} or linearized and discretized stochastic nonlinear systems~\cite{1469949}, finding the solution to the stochastic nonlinear state feedback $\mathcal{H}_{\infty}$ optimal control problem is not trivial in general.

The CV-STEM addresses this issue by numerically sampling an optimal contraction metric and CLF that greedily minimize an upper bound of the steady-state mean squared tracking error of It\^{o} stochastic nonlinear system trajectories. We select this as an objective function, instead of integral objective functions which often appear in optimal control problems, as it gives us an exact convex optimization-based control synthesis algorithm.  Also, since the problem has the SDRI as its constraint, the CV-STEM control is robust against both deterministic and stochastic disturbances and ensures that the tracking error is exponentially bounded for all time. We remark that this approach is not intended to supersede but to be utilized on top of existing methodologies on constructing desired control inputs using stochastic nonlinear optimal control techniques~\cite{doi:10.1137/0328054,nla.cat-vn712853,Bertsekas:2000:DPO:517430,1100238,194311}, as this is a type of feedback control scheme. In particular, stochastic model predictive control~\cite{7740982,6858851} with guaranteed stability~\cite{MAHMOOD20122271,7526514} \textit{assumes} the existence of a stochastic CLF, whilst our approach explicitly \textit{constructs} an optimal CLF which could be used for the stochastic CLF with some modifications on the non-vanishing error term in our formulation.


The tool we use for analyzing incremental stability~\cite{989067} in this paper is contraction analysis~\cite{contraction,Wang:2005:PCA:2733752.2733964,1428824}, where its stochastic version is derived in~\cite{stochastic_contraction,observer}. Contraction analysis for discrete-time and hybrid systems is provided in~\cite{contraction,hybrid,hybrid2} and its stochastic counterpart is investigated in~\cite{4795665} with respect to a state-independent metric. In this paper, we describe discrete-time incremental contraction analysis with respect to a state- and time-dependent metric. Since the differential (virtual) dynamics of $\delta x$ used in contraction analysis is a Linear Time-Varying (LTV) system, global exponential stability can be studied using a quadratic Lyapunov function of $\delta x$, $V=\delta x^{\top} M(x,t)\delta x$~\cite{contraction}, as opposed to the Lyapunov technique where $V$ could be any function of $x$. Therefore, designing $V$ reduces to finding a positive definite metric $M(x,t)$~\cite{AYLWARD20082163,ncm,nscm},  which enables the aforementioned convex optimization-based control of It\^{o} stochastic nonlinear systems.
\subsection{Paper Organization}
The rest of this paper is organized as follows. Section~\ref{preliminaries} introduces stochastic incremental contraction analysis and presents its discrete-time version with a state- and time-dependent metric. In Sec.~\ref{controller_design}, the CV-STEM control for It\^{o} stochastic nonlinear systems is presented and its stability is analyzed using contraction analysis. In Sec.~\ref{lagrangian_systems}, this approach is extended to the control of stochastic Lagrangian systems. Section~\ref{variations} elucidates several extensions of the CV-STEM control synthesis. The aforementioned two simulation examples are reported in Sec.~\ref{simulation}. Section~\ref{conclusion} concludes the paper.
\subsection{Notation}
\label{notation}
For a vector $x \in \mathbb{R}^n$ and a matrix $A \in \mathbb{R}^{n \times m}$, we let $\|x\|$, $\delta x$, $\partial_{\mu} x$, $\|A\|$, $\|A\|_F$, $\image(A)$, $\kernel(A)$, $A^{+}$, and $\kappa(A)$ denote the Euclidean norm, infinitesimal variation of $x$, partial derivative of $x$ with respect to $\mu$, induced 2-norm, Frobenius norm, image of $A$, kernel of $A$, Moore--Penrose inverse, and condition number, respectively. For a square matrix $A$, we use the notation $\lambda_{\min}(A)$ and $\lambda_{\max}(A)$ for the minimum and maximum eigenvalues of $A$, $\trace(A)$ for the trace of $A$, $A \succ 0$, $A \succeq 0$, $A \prec 0$, and $A \preceq 0$ for the positive definite, positive semi-definite, negative definite, negative semi-definite matrices, respectively, and $\sym(A) = (A+A^{\top})/2$. For a vector $x \in \mathbb{R}^n$ and a positive definite matrix $A \in \mathbb{R}^{n \times n}$, we denote a norm $\sqrt{x^{\top} Ax}$ as $\|x\|_A$. Also, $I \in \mathbb{R}^{n \times n}$ represents the identity matrix, $E[\cdot]$ denotes the expected value operator, and $E_{\xi}[\cdot]$ denotes the conditional expected value operator when $\xi$ is given.
The $\mathcal{L}_p$ norm in the extended space $\mathcal{L}_{pe}$, $p \in [1,\infty]$,  is defined as
$\|(y)_{\tau}\|_{\mathcal{L}_p} = \left(\int_0^\tau \|y(t)\|^p\right)^{{1}/{p}} < \infty$ for $p\in[1,\infty)$ and $\|(y)_{\tau}\|_{\mathcal{L}_{\infty}} = \sup_{t\geq 0}\|(y(t))_{\tau}\| < \infty$ for $p =\infty$,
where $(y(t))_{\tau}$ is a truncation of $y(t)$, \ie, $(y(t))_{\tau} = 0$ for $t > \tau$ and $(y(t))_{\tau} = y(t)$ for $0 \leq t \leq \tau$ with $\tau \in [0,\infty)$.
\section{Stochastic Incremental Stability via Contraction Analysis}
\label{preliminaries}
We summarize contraction analysis that will be used for stability analysis in the subsequent sections. This allows us to utilize approaches for LTV systems theory, yielding a convex optimization-based framework for optimal Lyapunov function construction in Sec.~\ref{controller_design} and~\ref{lagrangian_systems}.

We also present new theorems for analyzing stochastic incremental stability of discrete-time nonlinear systems with respect to a state- and time-dependent Riemannian metric, along with its explicit connection to contraction analysis of continuous-time systems.
\subsection{Continuous-time Dynamical Systems}
Consider the following continuous-time nonlinear non-autonomous system and its virtual dynamics:
\begin{align}
\label{eq1}
\dot{x} = f(x,t), \ \ 
\delta \dot{x} = \frac{\partial f(x,t)}{\partial x}\delta x
\end{align}
where $t\in \mathbb{R}_{\geq0}$, $x:\mathbb{R}_{\geq0} \to \mathbb{R}^{n}$, and $f:\mathbb{R}^n\times\mathbb{R}_{\geq0} \to \mathbb{R}^{n}$. Incremental stability~\cite{989067} is defined as stability of system trajectories with respect to each other by means of differential (virtual) dynamics. Contraction theory is used to study incremental stability with exponential convergence.
\begin{lemma}
\label{dcont}
The system (\ref{eq1}) is contracting (\ie{} all the solution trajectories exponentially converge to a single trajectory globally from any initial condition), if there exists a uniformly positive definite metric $M(x,t) = \Theta(x,t)^{\top}\Theta(x,t)$, $M(x,t) \succ 0,~\forall  x,t$, with a smooth coordinate transformation of the virtual displacement $\delta z = \Theta(x,t) \delta x$, such that
\begin{align}
\label{deterministic_contraction}
&\dot{M}(x,t)+2\sym\left(M(x,t)\frac{\partial f}{\partial x}\right) \preceq -2\gamma_c M(x,t),~\forall x,t 
\end{align}
where $\gamma_c > 0$.
If the system (\ref{eq1}) is contracting, then we have $\|\delta z(t)\|= \|\Theta(x,t) \delta x(t)\|  \leq \|\delta z(0)\|e^{-\gamma_c t}$.
\end{lemma}
\begin{IEEEproof}
See~\cite{contraction}.
\end{IEEEproof}
Next, consider the nonlinear system (\ref{eq1}) with stochastic perturbation given by the It\^{o} stochastic differential equation
\begin{align}
\label{stochastic_dynamics}
dx =& f(x,t)dt + G(x,t)d\mathscr{W},~x(0) = x_0
\end{align}
where $G:\mathbb{R}^n\times\mathbb{R}_{\geq0} \to \mathbb{R}^{n\times d}$ is a matrix-valued function, $\mathscr{W}(t)$ is a $d$-dimensional Wiener process, and $x_0$ is a random variable independent of $\mathscr{W}(t)$~\cite{arnold_SDE}.
In this paper, we assume that $\exists L_1 > 0$ \st{} $\|f(x_1,t)-f(x_2,t)\|+\|G(x_1,t)-G(x_2,t)\|_F \leq L_1 \|x_1-x_2\|$, $\forall t\in\mathbb{R}_{\geq0}$ and $\forall x_1,x_2 \in \mathbb{R}^n$, and $\exists L_2 > 0$ \st{} $\|f(x_1,t)\|^2+\|G(x_1,t)\|_F^2 \leq L_2 (1+\|x_1\|^2)$, $\forall t\in\mathbb{R}_{\geq0}$ and $\forall x_1 \in \mathbb{R}^n$, for the sake of existence and uniqueness of the solution to (\ref{stochastic_dynamics}).
Now, consider the following two systems with trajectories $\xi_1(t)$ and $\xi_2(t)$ driven by two independent Wiener processes $\mathscr{W}_1(t)$ and $\mathscr{W}_2(t)$:
\begin{align}
\label{eq4}
d\xi =& \begin{bmatrix} f(\xi_1,t) \\ f(\xi_2,t) \end{bmatrix} dt + \begin{bmatrix} G_1(\xi_1,t) & 0 \\0 & G_2(\xi_2,t) \end{bmatrix}\begin{bmatrix} d\mathscr{W}_1 \\ d\mathscr{W}_2 \end{bmatrix}
\end{align}
where $\xi(t) = [\xi_1(t)^{\top},\xi_2(t)^{\top}]^{\top}\in\mathbb{R}^{2n}$.
The following theorem analyzes stochastic incremental stability of the two trajectories $\xi_1(t)$ and $\xi_2(t)$ with respect to each other in the presence of stochastic noise. The trajectories of (\ref{stochastic_dynamics}) are parameterized as $x(0,t) = \xi_1$ and $x(1,t) = \xi_2$. Also, we define $G(x,t)$ as $G(x(0,t),t) = G_1(\xi_1,t)$ and $G(x(1,t),t) = G_2(\xi_2,t)$.
\begin{theorem}
\label{sic}
Suppose that there exist bounded positive constants $\underline{m}$, $\overline{m}$, $g_1$, $g_2$, $\overline{m}_{x}$, and $\overline{m}_{x^2}$ \st{} $\underline{m} \leq \|M(x,t)\| \leq \overline{m}$, $\|G_1(x,t)\|_F \leq g_1$, $\|G_2(x,t)\|_F \leq g_2$, $\|\partial (M_{ij})/\partial x\| \leq \overline{m}_{x}$, and $\left\|{\partial^2(M_{ij})}/{\partial x^2}\right\| \leq \overline{m}_{x^2},~\forall x,t$. 
Suppose also that (\ref{deterministic_contraction}) holds (\ie{}, the deterministic system (\ref{eq1}) is contracting). Consider the generalized squared length with respect to a Riemannian metric $M(x(\mu,t),t)$ defined by
\begin{align}
V(x,\partial_{\mu} x,t) = \int_0^1\frac{\partial x}{\partial \mu}^{\top}M(x(\mu,t),t)\frac{\partial x}{\partial \mu}d\mu
\end{align}
\st{} $V(x,\partial_{\mu} x,t) \geq \underline{m}\|\xi_1-\xi_2\|^2$. Then we have
\begin{align}
\label{lsic}
    \mathscr{L}V \leq -2\gamma_1V+\underline{m}C_c
\end{align}
for $\gamma_1 = \gamma_c-((g_1^2+g_2^2)/2\underline{m})(\varepsilon_c\overline{m}_{x}+\overline{m}_{x^2}/2)$ and $C_c=(\overline{m}/\underline{m}+\overline{m}_{x}/(\varepsilon_c\underline{m}))(g_1^2+g_2^2)$, where $\mathscr{L}$ is an infinitesimal differential generator~\cite{observer}, $\gamma_c$ is the contraction rate for the deterministic system (\ref{eq1}), and $\varepsilon_c>0$ is an arbitrary constant. Further, if we have $\gamma_1 > 0$, (\ref{lsic}) implies that the mean squared distance between the two trajectories of (\ref{eq4}), whose initial conditions given by a probability distribution $p(a_0,b_0)$ that are independent of $\mathscr{W}_1(t)$ and $\mathscr{W}_2(t)$, is exponentially bounded as follows:
\begin{align}
\label{boundnewsto}
E\left[\|\xi_1(t)-\xi_2(t)\|^2\right] \leq \frac{C_c}{2\gamma_1}+\frac{E[V(x(0),\partial_{\mu} x(0),0)]e^{-2\gamma_1 t}}{\underline{m}}.
\end{align}
\end{theorem}
\begin{IEEEproof}
Using the property $\trace(AB) \leq \|A\|\trace(B)$ for $A,B \succeq 0$, we have
$\trace(G_i(\xi_i,t)^{\top}M(\xi_i,t)G_i(\xi_i,t)) \leq \overline{m}g_i^2$. Therefore, computing $\mathscr{L}V$ as in the proof given in Lemma~2 of~\cite{observer} yields (\ref{lsic}). Taking expectation on both sides of (\ref{lsic}) along with Dynkin's formula~\cite[pp. 10]{nla.cat-vn712853} completes the derivation of (\ref{boundnewsto}).
\end{IEEEproof}
\begin{remark}
\label{epsilon_remark}
The contraction rate $\gamma_1$ and uncertainty bound $C_c$ depend on the choice of an arbitrary constant $\varepsilon_c$. One way to select $\varepsilon_c$ is to solve $dF/d\varepsilon_c = 0$ with $F(\varepsilon_c) = C_c/(2\gamma_1)$, whose solution minimizes the steady-state bound $F(\varepsilon_c)$ with the constraint $\gamma_1 >0$~\cite{observer}. Line search algorithms could also be used to select their optimal values~\cite{ncm,nscm}. We will utilize the fact that $C_c$ is a function of $\overline{m}/\underline{m}$ to facilitate the convex optimization-based control synthesis in Sec.~\ref{controller_design} and \ref{lagrangian_systems}.
\end{remark}
\subsection{Main Result 1: Connection between Continuous and Discrete Stochastic Incremental Contraction Analysis}
We establish a similar result to Lemma~\ref{dcont} for the following discrete-time nonlinear system and its virtual dynamics:
\begin{align}
\label{disc1} 
x_{k+1} = f_k(x_k,k), \ \
\delta x_{k+1} = \frac{\partial f_k(x_k,k)}{\partial x_k}\delta x_k
\end{align}
where $x_k \in \mathbb{R}^n$ and $f_k:\mathbb{R}^n\times\mathbb{N} \to \mathbb{R}^{n}$.
\begin{lemma}
The system (\ref{disc1}) is contracting if there exists a uniformly positive definite metric $M_k(x_k,k) = \Theta_k(x_k,k)^{\top}\Theta_k(x_k,k)$, $M_k(x_k,k) \succ 0,~\forall x_k,k$, with a smooth coordinate transformation of the virtual displacement $\delta z_k = \Theta_k(x_k,k) \delta x_k$ \st{}
\begin{align}
\label{discrete_deterministic_contraction}
&\frac{\partial f_k}{\partial x_k}^{\top} M_{k+1}(x_{k+1},k+1) \frac{\partial f_k}{\partial x_k} \preceq (1-\gamma_d) M_k(x_{k},k),~\forall x_k,k
\end{align}
where $\gamma_d \in (0,1)$.
If the system (\ref{disc1}) is contracting, then we have $\|\delta z_k\| = \|\Theta_k(x_k,k) \delta x_k\| \leq \|\delta z_0\|(1-\gamma_d)^{\frac{k}{2}}$.
\end{lemma}
\begin{IEEEproof}
See~\cite{contraction,hybrid2}.
\end{IEEEproof}
We now present a discrete-time version of Theorem~\ref{sic}, which can be extensively used for proving stability of discrete-time and hybrid stochastic nonlinear systems, along with known results for deterministic systems~\cite{hybrid,hybrid2}. Consider the discrete-time nonlinear system (\ref{disc1}) with stochastic perturbation modeled by the stochastic difference equation
\begin{align}
x_{k+1} =& f_k(x_k,k) + G_k(x_k,k)w_k \label{disc3}
\end{align}
where $G_k:\mathbb{R}^n\times\mathbb{N} \to \mathbb{R}^{n\times d}$ is a matrix-valued function and $w_k$ is a $d$-dimensional sequence of zero mean uncorrelated normalized Gaussian random variables.
Consider the following two systems with trajectories $\xi_{1,k}$ and $\xi_{2,k}$ driven by two independent stochastic perturbation $w_{1,k}$ and $w_{2,k}$:
\begin{align}
\label{disc4}
\xi_{k+1} =& \begin{bmatrix} f_k(\xi_{1,k},k) \\ f_k(\xi_{2,k},k) \end{bmatrix} + \begin{bmatrix} G_{1,k}(\xi_{1,k},k) & 0 \\0 & G_{2,k}(\xi_{2,k},k) \end{bmatrix}\begin{bmatrix} w_{1,k} \\ w_{2,k} \end{bmatrix}
\end{align}
where $\xi_k = [\xi_{1,k}^{\top},\xi_{2,k}^{\top}]^{\top}\in\mathbb{R}^{2n}$.
The following theorem analyzes stochastic incremental stability for discrete-time nonlinear systems, but we remark that this is different from~\cite{old_observer,4795665} in that the stability is studied in a differential sense and its Riemannian metric is state- and time-dependent. We parameterize $x_k$ and $G_k$ in (\ref{disc3}) as $x_k(\mu = 0) = \xi_{1,k}$, $x_k(\mu = 1) = \xi_{2,k}$, $G_k(x_k(\mu = 0),k) = G_{1,k}(\xi_{1,k},k)$, and $G_k(x_k(\mu = 1),k) = G_{2,k}(\xi_{2,k},k)$.
\begin{theorem}
\label{disc_sic}
Suppose that the system (\ref{disc4}) has the following bounds, $\underline{m}I \preceq M_k(x_k,k) \preceq \overline{m}I,~\forall x_k,k$, $\|G_{1,k}(\xi_{1,k},k)\|_F \leq g_{1d}$, and $\|G_{2,k}(\xi_{2,k},k)\|_F \leq g_{2d},~\forall \xi_{1,k},\xi_{2,k},k$, where $\overline{m}$, $g_{1d}$, and $g_{2d}$ are bounded positive constants.
Suppose also that (\ref{discrete_deterministic_contraction}) holds for the discrete-time deterministic system (\ref{disc1}) and there exists $\gamma_2 \in (0,1)$ \st{} $\gamma_2 \leq 1- (\overline{m}/\underline{m})(1-\gamma_d)$, where $\gamma_d$ is the contraction rate of (\ref{disc1}). Consider the generalized squared length with respect to a Riemannian metric $M_k(x_k(\mu),k)$ defined as 
\begin{align}
\label{eq:dts_Vdef}
V_k(x_k,\partial_{\mu} x_k,k) = \int_0^1\frac{\partial x_k}{\partial \mu}^{\top}M_k(x_k(\mu),k)\frac{\partial x_k}{\partial \mu}d\mu
\end{align} \st{} $V_k(x_k,\partial_{\mu} x_k,k) \geq \underline{m}\|\xi_{1,k}-\xi_{2,k}\|_2^2$. Then the mean squared distance between the two trajectories of the system (\ref{disc4}) is bounded as follows:
\begin{align}
\label{disc_msd}
E\left[\|\xi_{1,k}-\xi_{2,k}\|^2\right] \leq& \frac{1-\tilde{\gamma}_d^k}{1-\tilde{\gamma}_d}C_d+\frac{\tilde{\gamma}_d^k}{\underline{m}} E[V_{0}(x_0,\partial_{\mu} x_0,0)].
\end{align}
where $C_d = (\overline{m}/\underline{m})(g_{1d}^2+g_{2d}^2)$ and $\tilde{\gamma}_d = 1-\gamma_2 \in (0,1)$.
\end{theorem}
\begin{IEEEproof}
Consider a Lyapunov-like function $V_{k}$ in (\ref{eq:dts_Vdef}), where we use $V_{k} = V_k(x_k,\partial_{\mu} x_k,k)$ and $M_{k} = M_k(x_{k},k)$ for notational simplicity. Using the bounds along with (\ref{discrete_deterministic_contraction}) and (\ref{disc3}), we have, for $\ell\in\mathbb{N}$, that
\begin{align}
\label{tochu}
&V_{{\ell}+1} \leq \overline{m}\int_0^1\left\|\frac{\partial f_{\ell}}{\partial x_{\ell}}\frac{\partial x_{\ell}}{\partial \mu}+\frac{\partial G_{\ell}}{\partial \mu}w_{\ell}\right\|^2d\mu  \\
&\leq \frac{\overline{m}}{\underline{m}}(1-\gamma_d)\int_0^1\frac{\partial x_{{\ell}}}{\partial \mu}^{\top}M_{{\ell}}\frac{\partial x_{{\ell}}}{\partial \mu}d\mu \nonumber \\
& +\overline{m}\int_0^1 \left(2\frac{\partial x_{\ell}}{\partial \mu}^{\top}\frac{\partial f_{\ell}}{\partial x_{\ell}}^{\top}\frac{\partial G_{\ell}}{\partial \mu}w_{\ell}+w_{\ell}^{\top}\frac{\partial G_{\ell}}{\partial \mu}^{\top}\frac{\partial G_{\ell}}{\partial \mu}w_{\ell}\right)d\mu\nonumber
\end{align}
where $f_{\ell}=f_{\ell}(x_{\ell},{\ell})$ and $G_{{\ell}}=G_{{\ell}}(x_{\ell},{\ell})$. Taking the conditional expected value of $(\ref{tochu})$ when $x_{\ell}$, $\partial_{\mu} x_{\ell}$, and ${\ell}$ are given, we have that (see also: Theorem~2 of~\cite{old_observer})
\begin{align}
\label{disc_con_real}
&E_{\zeta_{\ell}}[V_{{\ell}+1}] \leq \gamma_m V_{\ell}+\overline{m}E_{\zeta_{\ell}}\left[\int^1_0w_{\ell}^{\top}\frac{\partial G_{\ell}}{\partial \mu}^{\top}\frac{\partial G_{\ell}}{\partial \mu}w_{\ell}d\mu\right] \nonumber \\
&\leq \gamma_m V_{\ell}+\sum_{i=1,2}\overline{m}E_{\zeta_{\ell}}\left[\trace\left(w_{i,{\ell}}w_{i,{\ell}}^{\top}G_{i,{\ell}}^{\top}G_{i,{\ell}}\right)\right] \nonumber \\
&\leq \gamma_m V_{\ell}+\overline{m}\sum_{i=1,2}\trace\left(G_{i,{\ell}}^{\top}G_{i,{\ell}}\right) \leq \tilde{\gamma}_d V_{\ell}+\underline{m}C_d.
\end{align}
where $\gamma_m = \overline{m}/\underline{m}(1-\gamma_d)$, and $x_{\ell}$, $\partial_{\mu} x_{\ell}$, and ${\ell}$ are denoted as $\zeta_{\ell}$. Here, we used the condition: $\exists \gamma_2\in(0,1)$ \st{} $\gamma_m \leq 1-\gamma_2=\tilde{\gamma}_d$. Taking expectation over $\zeta_{\ell-1}$ in (\ref{disc_con_real}) with the tower rule $E_{\zeta_{{\ell}-1}}[V_{{\ell}+1}] = E_{\zeta_{{\ell}-1}}[E_{\zeta_{{\ell}}}[V_{{\ell}+1}]]$ gives us that
\begin{align}
E_{\zeta_{{\ell}-1}}[V_{{\ell}+1}] \leq \tilde{\gamma}_d^2V_{{\ell}-1}+\underline{m}C_d+\underline{m}C_d\tilde{\gamma}_d
\end{align}
where $\tilde{\gamma}_d = 1-\gamma_2$. Continuing this operation with the relation $\underline{m}E_{\zeta_{0}}\left[\|\xi_{1,\ell+1}-\xi_{2,{\ell+1}}\|^2\right] \leq E_{\zeta_{0}}\left[V_{\ell+1}\right]$  yields
\begin{align}
E_{\zeta_{0}}\left[\|\xi_{1,k}-\xi_{2,k}\|^2\right]-\frac{\tilde{\gamma}_d^k}{\underline{m}} V_{0} \leq C_d\sum_{i=0}^{k-1}\tilde{\gamma}_d^i = \frac{1-\tilde{\gamma}_d^k}{1-\tilde{\gamma}_d}C_d \nonumber
\end{align}
where $k = \ell+1$. Taking expectation over $\zeta_{0}$ and rearranging terms result in (\ref{disc_msd}).
\end{IEEEproof}

Let us now consider the case where the time interval $\Delta t = t_{k+1} -t_k$ is sufficiently small, \ie, $\Delta t \gg (\Delta t)^2$. Then the continuous-time stochastic system (\ref{stochastic_dynamics}) can be discretized as
\begin{align}
\label{stochastic_disc_eq}
x_{k+1} &= x_{k}+\int_{t_k}^{t_{k+1}}f(x(t),t)dt+G(x(t),t)d\mathscr{W}(t) \nonumber \\
&\simeq x_{k}+f(x_{k},t_k)\Delta t+G(x_{k},t_k)\Delta \mathscr{W}_k
\end{align}
where $x_{k} = x(t_k)$, $\Delta \mathscr{W}_k = \sqrt{\Delta t}w_k$, and $w_k$ is a $d$-dimensional sequence of zero mean uncorrelated normalized Gaussian random variables. When $\Delta t \gg (\Delta t)^2$, $f_k(x_k,k)$ and $G_k(x_k,k)$ in (\ref{disc3}) can be approximated as $f_k(x_k,k) = x_k+f(x_k,t_k)\Delta t$ and $G_k(x_k,k) = \sqrt{\Delta t} G(x_k,t_k)$. In this situation, we have the following theorem that connects stochastic incremental stability of discrete-time systems with that of continuous-time systems.
\begin{theorem}
\label{disc_cont_thm}
Suppose that (\ref{disc_con_real}) in Theorem~\ref{disc_sic} holds with $\tilde{\gamma}_d=1-\gamma_2 \in (0,1)$. Then the expected value of $V_{k+1}$ up to first order in $\Delta t$ is given as $E_{\zeta_k}[V_{k+1}] = V_k+\Delta t\mathscr{L}V_k$, where $\mathscr{L}$ is an infinitesimal differential generator. Furthermore, the following inequality holds:
\begin{align}
\label{disc_cont_rel}
\mathscr{L}V_k(x_k,\partial_{\mu} x_k,t_k) \leq -\frac{\gamma_2}{\Delta t}V_k(x_k,\partial_{\mu} x_k,t_k)+\underline{m}\tilde{C}_c
\end{align}
where $\tilde{C}_c$ is a positive constant given as
\begin{align}
\label{Cceq}
\tilde{C}_c = \frac{C_d}{\Delta t} = \frac{\overline{m}}{\underline{m}\Delta t}(g_{1d}^2+g_{2d}^2) = \frac{\overline{m}}{\underline{m}}(g_{1}^2+g_{2}^2)    
\end{align}
with $g_1$ and $g_2$ defined in Theorem~\ref{sic}.
\end{theorem}
\begin{IEEEproof}
$M_{k+1}$ up to first order in $\Delta t$ is written as 
\begin{align}
\label{M_kp1}
&M_{k+1} = \frac{\partial M_k}{\partial t_k}\Delta t+\sum_{i=1}^n\frac{\partial M_k}{\partial (x_{k})_i}(f_{c,k}\Delta t+G_{c,k}\Delta \mathscr{W}_k)_i \\
&+\frac{1}{2}\sum_{i=1}^n\sum_{j=1}^n\frac{\partial^2 M_k}{\partial (x_{k})_i\partial (x_{k})_j}(G_{c,k}\Delta \mathscr{W}_k)_i(G_{c,k}\Delta \mathscr{W}_k)_j+M_k \nonumber
\end{align}
where $f_{c,k}$ and $G_{c,k}$ are defined as $f_{c,k} = f(x_k,t_k)$ and $G_{c,k} = G(x_k,t_k)$ for notational simplicity. The subscripts $i$ and $j$ denote the $i$th and $j$th element of the corresponding vectors.
Similarly, ${\partial x_{k+1}}/{\partial \mu}$ up to first order in $\Delta t$ can be computed as 
\begin{align}
\label{x_kp1_mu}
\frac{\partial x_{k+1}}{\partial \mu} = \frac{\partial x_{k}}{\partial \mu}+\frac{\partial f_{c,k}}{\partial x_k}\frac{\partial x_{k}}{\partial \mu}\Delta t+\frac{\partial G_{c,k}}{\partial \mu}\Delta \mathscr{W}_k.
\end{align}
Substituting (\ref{M_kp1}) and (\ref{x_kp1_mu}) into $E_{\zeta_k}[V_{k+1}]$ yields
\begin{align}
E_{\zeta_k}[V_{k+1}] &= E_{\zeta_k}\left[\int_0^1\frac{\partial x_{k+1}}{\partial \mu}^{\top}M_{k+1}\frac{\partial x_{k+1}}{\partial \mu}d\mu\right] \nonumber \\
&= V_k+(dV_{d,k}+dV_{s,k})\Delta t+\mathcal{O}(\Delta t^{{3}/{2}})
\end{align}
where $dV_{d,k}$ and $dV_{s,k}$ are given by
\begin{align}
dV_{d,k} =& \int_0^1\frac{\partial x_k}{\partial \mu}^{\top}\left(\frac{\partial f_{c,k}}{\partial x_k}^{\top}M_k+\dot{M}_k+M_k\frac{\partial f_{c,k}}{\partial x_k}\right)\frac{\partial x_k}{\partial \mu}d\mu
\end{align}
with $\dot{M}_k = {\partial M_k}/{\partial t_k}+\sum_{i=1}^n({\partial M_k}/{\partial (x_{k})_i})f_{c,k}$ and
\begin{align}
dV_{s,k} =& \int_0^1\left[\sum_{i=1}^n\sum_{j=1}^n(M_k)_{ij}\left(\frac{\partial G_{c,k}}{\partial \mu}\frac{\partial G_{c,k}}{\partial \mu}^{\top}\right)_{ij} \nonumber \right. \\
& \left. + 2\frac{\partial (M_k)_i}{\partial (x_{k})_j}\frac{\partial x_k}{\partial \mu}\left(G_{c,k}\frac{\partial G_{c,k}}{\partial \mu}^{\top}\right)_{ij} \nonumber \right. \\
& \left.+ \frac{1}{2}\frac{\partial x_k}{\partial \mu}^{\top}\frac{\partial^2 M_k}{\partial (x_{k})_i\partial (x_{k})_j}\frac{\partial x_k}{\partial \mu} (G_{c,k}G_{c,k}^{\top})_{ij}\right] d\mu.
\end{align}
We note that the properties of $w_k$ as a $d$-dimensional sequence of zero mean uncorrelated normalized Gaussian random variables are used to derive these relations.
Since $dV_{d,k}+dV_{s,k} = \mathscr{L}V_k$ where $\mathscr{L}$ is the infinitesimal differential generator, we have $E_{\zeta_k}[V_{k+1}] = V_k+\Delta t\mathscr{L}V_k$. Thus, the condition $E_{\zeta_k}[V_{k+1}] \leq (1-\gamma_2)V_k+\underline{m}C_d$ given by (\ref{disc_con_real}) in Theorem~\ref{disc_sic} reduces to the following inequality:
\begin{align}
\label{last_disc_cont}
\mathscr{L}V_k(x_k,\partial_{\mu} x_k,t_k) \leq -\frac{\gamma_2}{\Delta t}V_k(x_k,\partial_{\mu} x_k,t_k)+\underline{m}\frac{C_d}{\Delta t}.
\end{align}
Finally, (\ref{last_disc_cont}) with the relations $\tilde{C}_c = {C_d}/{\Delta t}$ and $G_k(x_k,k) = \sqrt{\Delta t} G(x_k,t_k)$ results in (\ref{disc_cont_rel}) and (\ref{Cceq}).
\end{IEEEproof}
\begin{remark}
The positive constant $\tilde{C}_c$ is equal to the positive constant $C_c$ in Theorem~\ref{sic} when $\overline{m}_{x}=0$. This is due to the fact that we used an upper bound of $\|M_k\|$ when obtaining the first line of (\ref{tochu}) in Theorem~\ref{disc_sic}.
\end{remark}

In practical control applications, we use the same control input at $t = t_k$ for a finite time interval $t \in [t_k,t_{t+1})$. Theorems~\ref{sic} and \ref{disc_cont_thm} indicate that if $\Delta t$ is sufficiently small, a discrete-time stochastic controller can be viewed as a continuous-time counterpart with contraction rate $2\gamma_1 = \gamma_2/\Delta t$. We will illustrate how to select the sampling period $\Delta t$ large enough without deteriorating the CV-STEM control performance in Sec.~\ref{simulation}. Also, the steady-state mean squared tracking error for both discrete and continuous cases can be expressed as a function of the condition number of the metric $M(x,t)$, which is useful in designing convex optimization-based control synthesis as shall be seen in Sec.~\ref{controller_design} and~\ref{lagrangian_systems}.
\section{Main Result 2: CV-STEM Control with Stability and Optimization}
\label{controller_design}
This section presents the CV-STEM control for general input-affine nonlinear stochastic systems, incremental stability of which is analyzed using contraction theory given in Theorems~\ref{sic}~and~\ref{disc_cont_thm}. Since the differential dynamics of $\delta x$ used in contraction analysis can be viewed as an LTV system, we can use an optimal differential Lyapunov function of the form $\delta x^{\top} M(x,t)\delta x$ without loss of generality~\cite{contraction}, thereby finding $M(x,t)$ via convex optimization. We note that this is not for finding an optimal control trajectory and input, which can be used as a desired trajectory in the present control design. 

In Sec.~\ref{optimal_controller_design}, we present a convex optimization problem for finding the optimal contraction metric for the CV-STEM control, which greedily minimizes an upper bound of the steady-state mean squared tracking error of It\^{o} stochastic nonlinear system trajectories. It is shown that this problem is equivalent to the original non-convex optimization problem of minimizing the upper bound.
\subsection{Problem Formulation}
Consider the following It\^{o} stochastic nonlinear systems with a control input $u$, perturbed by a $d$-dimensional Wiener process $\mathscr{W}(t)$:
\begin{align}
\label{original_dynamics}
dx &= f(x,t)dt + B(x,t)udt + G_u(x,t)d\mathscr{W} \nonumber \\
dx_d &= f(x_d,t)dt + B(x_d,t)u_ddt.
 \end{align}
where $u:\mathbb{R}_{\geq0} \to \mathbb{R}^{m}$, $B:\mathbb{R}^n\times\mathbb{R}_{\geq0} \to \mathbb{R}^{n\times m}$, $G_u:\mathbb{R}^n\times\mathbb{R}_{\geq0} \to \mathbb{R}^{n\times d}$, and $x_d:\mathbb{R}_{\geq0} \to \mathbb{R}^{n}$ and $u_d:\mathbb{R}_{\geq0} \to \mathbb{R}^{m}$ are the desired state and input, respectively. The dynamical system of the desired state is deterministic as $x_d$ and $u_d$ are assumed to be given.
\begin{remark}
Since $\dot{x}_d-f(x_d,t)\in \image B(x_d,t)$ holds for a feasible desired trajectory, $u_d$ can be obtained as
$u_d = B(x_d,t)^{+}(\dot{x}_d-f(x_d,t))$
where $(\cdot)^{+}$ denotes the Moore-Penrose inverse.
This is the unique least-squares solution (LSS) to $B(x_d,t)u_d = \dot{x}_d-f(x_d,t)$ when $\kernel B(x_d,t)=\{0\}$ and an LSS with the smallest Euclidean norm when $\kernel B(x_d,t)\neq\{0\}$. The desired input $u_d$ can also be found by solving an optimal control problem~\cite{citeulike:802300,doi:10.1137/0328054,nla.cat-vn712853,Bertsekas:2000:DPO:517430,1100238,194311,7740982,6858851,MAHMOOD20122271,7526514} and a general system with $\dot{x}=f(x,u)$ can be transformed into an input-affine form by treating $\dot{u}$ as another input.
\end{remark}

In the proceeding discussion, we assume that $f(x,t) = 0$ at $x = 0$ and that $f$ is a continuously differentiable function. This allows us to use the following lemma.
\begin{lemma}
\label{sdclemma}
Let $\Omega$ be the state set that is a bounded open subset of some Euclidean space \st{} $0 \in \Omega \subseteq \mathbb{R}^n$.
Under the assumptions $f(0) = 0$ and $f(x)$ is a continuously differentiable function of $x$ on  $\Omega$, there always exists at least one continuous nonlinear matrix-valued function $A(x)$ on $\Omega$ \st{} $f(x) = A(x)x$, where $A:\Omega \to \mathbb{R}^{n\times n}$ is found by mathematical factorization and is non-unique when $n > 1$.
\end{lemma}
\begin{IEEEproof}
See~\cite{survey_SDRE}.
\end{IEEEproof}
Using Lemma~\ref{sdclemma}, (\ref{original_dynamics}) is expressed as
\begin{align}
\label{sdc_dynamics}
dx &= A(\varrho,x,t)xdt + B(x,t)udt + G_u(x,t)d\mathscr{W} \nonumber \\
dx_d &= A(\varrho,x_d,t)x_ddt + B(x_d,t)u_ddt
\end{align}
where $\varrho=(\varrho_1,\cdots,\varrho_{s_1})$ are the coefficients of the convex combination of SDC parameterizations $A_i(x,t)$, \ie,
\begin{align}
A(\varrho,x,t)=\sum_{i=1}^{s_1}\varrho_iA_i(x,t).
\end{align}
Writing the system dynamics (\ref{original_dynamics}) in SDC form provides a design flexibility to mitigate effects of stochastic noise while verifying that the system is controllable as shall be seen later.
\subsection{Feedback Control Design}
We consider the following feedback control scheme (to be optimized in Sec.~\ref{optimal_controller_design}):
\begin{align}
u &= -K(x,t)(x-x_d)+u_d \nonumber \\
\label{controller}
&= -R(x,t)^{-1}B(x,t)^{\top}M(x,t)(x-x_d)+u_d
\end{align}
where $R(x,t) \succ 0$ is a weight matrix on the input $u$ and $M(x,t)$ is a positive definite matrix which satisfies the following matrix inequality for $\gamma > 0$:
\begin{align}
\label{riccati}
&\dot{M}(x,t)+2\sym(M(x,t)A(\varrho,x,t))+\gamma M^2(x,t) \nonumber \\
&-M(x,t)B(x,t)R(x,t)^{-1}B(x,t)^{\top}M(x,t) \preceq 0.
\end{align}
Define $A_{cl}(\varrho,y,t)$, $\Delta A(\varrho,y,t)$, and $\Delta B(y,t)$~\cite{Banks2007} as
\begin{align}
A_{cl}(\varrho,y,t) &= A(\varrho,y+x_d,t)-B(y+x_d,t)K(y+x_d,t) \nonumber \\
\Delta A(\varrho,y,t) &= A(\varrho,y+x_d,t)-A(\varrho,x_d,t)  \nonumber \\
\Delta B(y,t)  &= B(y+x_d,t)-B(x_d,t).
\end{align}
Substituting (\ref{controller}) into (\ref{sdc_dynamics}) yields
\begin{align}
\label{closed_loop_e}
de = f_v(e,t)dt+G_u(e+x_d,t)d\mathscr{W}
\end{align}
where $e = x-x_d$ and
\begin{align}
\label{def_virtual}
f_v(y,t) = A_{cl}(\varrho,e,t)y+\Delta A(\varrho,y,t) x_d+\Delta B(y,t) u_d. 
\end{align}
\begin{lemma}
\label{lemma_det_robustness_general}
Suppose that the deterministic system is perturbed as follows:
\begin{align}
\dot{x} = f(x,t) + B(x,t)(u+d).
\end{align}
If there exists a positive definite solution $M(x,t)$ to the inequality (\ref{riccati})
with $R(x,t) = S(x,t)^2 \succ 0$ and $S(x,t) \succ 0$, then the system with inputs $\mu_1 = S(x,t)d$, $\mu_2 = (\sqrt{{2}/{\gamma}})\Delta_d$ and an output $y = (\sqrt{{\gamma}/{2}})M(x,t)(x-x_d)$, where $\Delta_d = \Delta A x_d+\Delta B u_d$, is finite-gain $\mathcal{L}_2$ stable and its $\mathcal{L}_2$ gain is less than or equal to 1 for each input $\mu_1$ and $\mu_2$.
\end{lemma}
\begin{IEEEproof}
See Appendix \ref{proof_det_robustness}.
\end{IEEEproof}
\subsection{Incremental Stability Analysis}
As we discussed earlier in Sec.~\ref{preliminaries}, even when a control input at $t = t_k$ is applied during a finite time interval $t \in [t_k,t_{t+1})$, Theorem~\ref{disc_cont_thm} along with Theorem~\ref{disc_sic} guarantees that the discrete-time controller leads to an analogous result to the continuous-time case (\ref{controller}) if $\Delta t_k$ is sufficiently small. Thus, we perform stability analysis for continuous-time dynamical systems.
Let us define a deterministic virtual system of (\ref{sdc_dynamics}) as follows:
\begin{align}
\label{detvd}
\dot{y} =  f_v(y,t) = A_{cl}(\varrho,e,t)y+\Delta A(\varrho,y,t) x_d+\Delta B(y,t) u_d.
\end{align}
where (\ref{detvd}) has $y=e$ and $y=0$ as its particular solutions.
The virtual dynamics of (\ref{detvd}) is expressed as
\begin{align}
\delta\dot{y} = A_{cl}(\varrho,e,t)\delta y+\phi(\varrho,y,t)\delta y
\end{align}
where $\phi(\varrho,y,t) = {\partial \left(\Delta Ax_d + \Delta Bu_d\right)}/{\partial y}$.
Using $f_v(y,t)$, the virtual system of (\ref{closed_loop_e}) with respect $y$ is defined as
\begin{align}
\label{svd}
dy = f_v(y(\mu,t),t)dt + G(y(\mu,t),t)d\mathscr{W}
\end{align}
where $\mu \in [0,1]$ is introduced to parameterize the trajectories $y=e$ and $y=0$, \ie, $y(\mu=0,t) = e$, $y(\mu=1,t) = 0$, $G(y(0,t),t) = G_u(e+x_d,t)$, and $G(y(1,t),t) = 0_{n \times d}$. It can be seen that (\ref{svd}) has $y=e$ and $y=0$ as its particular solutions because we have
\begin{itemize}
\item (\ref{svd}) reduces to (\ref{closed_loop_e}) when $y=e$.
\item $f_v= \Delta A(\varrho,0,t) x_d+\Delta B(0,t) u_d=0$  and $G=0_{n \times d}$ when $y=0$.
\end{itemize}
Now we present the following theorem for exponential boundedness of the mean squared tracking error of system trajectories (\ref{sdc_dynamics}).
\begin{theorem}
\label{thmcontracting}
Suppose there exist bounded positive constants $\underline{m}$, $\overline{m}$, $\overline{m}_{x}$, $\overline{m}_{x^2}$, and $g_u$ \st{} $\underline{m} \leq \|M(x,t)\| \leq \overline{m}$, $\|\partial(m_{ij})/\partial x\| \leq \overline{m}_{x}$, $\left\|{\partial^2(m_{ij})}/{\partial x^2}\right\| \leq \overline{m}_{x^2}$, and $\|G_u(x,t)\|_F \leq g_u,~\forall x,t$ where $\underline{m} = \inf_{x,t}\lambda_{\min}(M(x,t))$, $\overline{m} = \sup_{x,t}\lambda_{\max}(M(x,t))$, and $m_{ij}$ is the ($i,j$) component of $M(x,t)$. Suppose also that there exists $\alpha > 0$ \st{}
\begin{align}
\label{alpha_cond}
&\gamma M^2+MBR^{-1}B^{\top}M-\phi^{\top}M-M\phi-2\alpha_g I \succeq 2\alpha M
\end{align}
where $2\alpha_g = g_u^2\left(\overline{m}_{x}\varepsilon+\overline{m}_{x^2}/2\right)$ with an arbitrary positive constant $\varepsilon$, and the arguments $\varrho$, $x$, and $t$ are dropped for notational simplicity.
If there exists a positive definite solution $M(x,t)$ to the inequalities (\ref{riccati}) and (\ref{alpha_cond}), then the mean squared distance between the trajectories of (\ref{sdc_dynamics}) under the feedback control (\ref{controller}) is exponentially bounded as follows:
\begin{align}
\label{stochasticmse}
E\left[\|x_d-x\|^2\right] \leq \frac{C}{2\alpha}+\frac{E[V(x(0),\partial_{\mu} y(0),0)]e^{-2\alpha t}}{\underline{m}}
\end{align}
where $V(x,\partial_{\mu} y,t) = \int_0^1I_V(x,\partial_{\mu} y,t)d\mu$ with
\begin{align}
\label{lyap_xy}
I_V(x,\partial_{\mu} y,t) = \frac{\partial y}{\partial \mu}^{\top}M(x,t)\frac{\partial y}{\partial \mu}
\end{align}
and $C = (\overline{m}/\underline{m})g_u^2+(\overline{m}_{x}g_u^2)/(\varepsilon\underline{m})$.
\end{theorem}
\begin{IEEEproof}
For notational simplicity, let $I_V = I_V(x,\partial_{\mu}y,t)$, $A = A(\varrho,x,t)$, $B = B(x,t)$, $R = R(x,t)$, $G = G(y,t)$, $M = M(x,t)$, and
$\phi = \phi(\varrho,y,t) = {\partial (\Delta Ax_d)}/{\partial y}+{\partial (\Delta Bu_d)}/{\partial y}$.
By using an infinitesimal differential generator $\mathscr{L}$, we obtain
\begin{align}
\label{l}
\mathscr{L}V =&\int_{0}^{1} \frac{\partial I_V}{\partial t}+\sum^n_{i=1}\left(\frac{\partial I_V}{\partial x_i}f_i+\frac{\partial I_V}{\partial (\partial_{\mu} y_i)}\frac{\partial f_v}{\partial y}\frac{\partial y}{\partial \mu}\right) \nonumber \\
&+\frac{1}{2}\sum^n_{i=1}\sum^n_{j=1}\Biggl[\frac{\partial^2 I_V}{\partial x_i\partial x_j}(G_u(x,t)G_u(x,t)^{\top})_{ij} \nonumber \\
& +2\frac{\partial^2 I_V}{\partial x_i\partial (\partial_{\mu} y_j)}\left(G_u(x,t)\frac{\partial G(y,t)}{\partial \mu}^{\top}\right)_{ij} \\
& +\frac{\partial^2 I_V}{\partial (\partial_{\mu} y_i)(\partial_{\mu} y_j)} \left(\frac{\partial G(y,t)}{\partial \mu}\frac{\partial G(y,t)}{\partial \mu}^{\top}\right)_{ij} \Biggr] d\mu \nonumber
\end{align}
where $f_i$ is the $i$th component of $f(x,t)$. Since we have 
\begin{align}
    \frac{\partial I_V}{\partial t}+\sum^{n}_{i=1}\frac{\partial I_V}{\partial x_i} f_i =& \frac{\partial y}{\partial \mu}^{\top}\dot{M}\frac{\partial y}{\partial \mu} \nonumber \\
    \sum^{n}_{i=1}\frac{\partial I_V}{\partial (\partial_{\mu} y_i)}\frac{\partial f_v}{\partial y}\frac{\partial y}{\partial \mu} =& 2\frac{\partial y}{\partial \mu}^{\top}\sym(M(A_{cl}(\rho,e,t)+\phi))\frac{\partial y}{\partial \mu} \nonumber
\end{align}
where (\ref{lyap_xy}) defines $I_V$, (\ref{l}) reduces to
\begin{align}
\label{lv_riccati0}
\mathscr{L}V
=& \int^1_0 \frac{\partial y}{\partial \mu}^{\top}(\dot{M}+A^{\top}M+MA-2MBR^{-1}B^{\top}M \nonumber \\
& +\phi^{\top}M+M\phi)\frac{\partial y}{\partial \mu}d\mu+V_2.
\end{align}
The computation of $V_2$ and its upper bound $\overline{V}_2 = 2\alpha_g\int_0^1\left\|{\partial y}/{\partial \mu}\right\|^2d\mu+\underline{m}C$ is given in Appendix \ref{computation_V2}. Substituting (\ref{riccati}) into (\ref{lv_riccati0}) yields
\begin{align}
\label{lv_riccati}
\mathscr{L}V
\leq&\int^1_0 \frac{\partial y}{\partial \mu}^{\top}(-\gamma M^2-MBR^{-1}B^{\top}M \\ \nonumber
&+\phi^{\top}M+M\phi)\frac{\partial y}{\partial \mu}d\mu+V_2.
\end{align}
Thus, using (\ref{alpha_cond}) and $V_2 \leq \overline{V}_2$, we have that
\begin{align}
\label{final_eqn}
\mathscr{L}V
\leq& -2\int_0^1\frac{\partial y}{\partial \mu}^{\top}(\alpha M+\alpha_g I)\frac{\partial y}{\partial \mu}d\mu \nonumber \\
&+2\alpha_g\int_0^1\left\|\frac{\partial y}{\partial \mu}\right\|^2d\mu+\underline{m}C \nonumber \\
=& -2\alpha V+\underline{m}C.
\end{align}
Theorem~\ref{sic} along with (\ref{final_eqn}) completes the derivation of (\ref{stochasticmse}).
\end{IEEEproof}
\begin{remark}
The Euclidean norm of the state vector has to be upper bounded by a constant~\cite{1470537,Banks2007} in order for (\ref{alpha_cond}) to have a positive definite solution and for $\|\phi\|$ to be bounded~\cite{1470537,observer}. This assumption is satisfied by many engineering applications~\cite{observer} and does not imply any assumption on the incremental stability of the proposed controller. Also, the result of Theorem~\ref{thmcontracting} does not imply the asymptotic almost-sure bounds as $V(x,\partial_{\mu} y,t)$ is not a supermartingale due to the non-vanishing term $\underline{m}C$ in (\ref{final_eqn}). Finite time bounds can be obtainable using the supermartingale inequality (see \cite[pp. 86]{nla.cat-vn712853},\cite{stochastic_contraction}).
\end{remark}
\subsection{Robustness against Stochastic and Deterministic Disturbances}
We also show that the tracking error has a finite $\mathcal{L}_2$ gain with respect to the noise and disturbances acting on the system, \ie{}, the proposed controller is robust against external deterministic and stochastic disturbances analogously to Lemma~\ref{lemma_det_robustness_general}.
Consider the following nonlinear system under these disturbances:
\begin{align}
\label{ds_dynamics}
dx = f(x,t)dt + B(x,t)udt + d(x,t)dt + G_u(x,t)d\mathscr{W}.
\end{align}
The virtual system is defined as
\begin{align}
\label{ds_vsys}
dy = f_v(y,t)dt + d_y(y,t)dt + G(y,t)d\mathscr{W}
\end{align}
where $d_y(e,t) = d(x,t)$ and $d_y(0,t) = 0$. Also, $f_v$ is defined in (\ref{def_virtual}) and $G$ is in (\ref{svd}).
One important example of these systems is a parametric uncertain system, where $d(x,t)$ is given as $d(x,t) = f_\mathrm{true}(x,t)-f(x,t)$ with $f_\mathrm{true}$ being the system with true parameter values. Thus, the following corollary allows us to apply adaptive control techniques including~\cite{lopez2019contraction,nakka2020chanceconstrained} on top of our method. In particular, it is shown in~\cite{lopez2019contraction} that we can use contraction metrics to estimate unknown parameters $\theta$ when $G_u(x,t) = 0$ and $d(x,t) = \Delta(x,t) \theta$ for a given $\Delta$.
\begin{corollary}
\label{robust_coro}
The controller (\ref{controller}) with the constraints (\ref{riccati}) and (\ref{alpha_cond}) is robust against external disturbances in (\ref{ds_dynamics}) and satisfies the following $\mathcal{L}_2$ norm bound on the tracking error $e$:
\begin{align}
\label{ds_robust}
E[\|(e)_{\tau}\|^2_{\mathcal{L}_2}] \leq \frac{E[\|e(0)\|_{M(0)}^2]+\frac{\overline{m}}{\varepsilon_1}E[\|(d)_{\tau}\|^2_{\mathcal{L}_2}]+C_m\tau}{2\alpha_1}
\end{align}
where $C_m = \underline{m}C$ and $\alpha_1 = \alpha\underline{m}-\varepsilon_1\overline{m}/2$ with some positive constant $\varepsilon_1$ that guarantees $\alpha_1 > 0$.
\end{corollary}
\begin{IEEEproof}
Using the controller (\ref{controller}) with (\ref{riccati}) and (\ref{alpha_cond}),
\begin{align}
\label{ditdis_tochu_general}
&\mathscr{L}V
\leq -2\alpha V+\underline{m}C+2\overline{m}\int_{0}^{1}\left\|\frac{\partial y}{\partial \mu}\right\|\left\|\frac{\partial d_y}{\partial \mu}\right\|d\mu \\
&\leq -\left(2\alpha\underline{m}-\varepsilon_1\overline{m}\right)\int_{0}^{1}\left\|\frac{\partial y}{\partial \mu}\right\|^2d\mu +C_m+\frac{\overline{m}}{\varepsilon_1}\left\|d(x,t)\right\|^2 \nonumber
\end{align}
where the inequality $2a'b' \leq \varepsilon_1^{-1}a'^2+\varepsilon_1 b'^2$ for scalars $a'$, $b'$ and $\varepsilon_1 > 0$ is used with $a' = \|\partial d_y/\partial \mu\|$ and $b' = \|\partial y/\partial \mu\|$.
Since $\varepsilon_1$ is arbitrary, let us select $\varepsilon_1$ \st{} $\alpha_1 = \alpha\underline{m}-\varepsilon_1\overline{m}/2 > 0$.
Applying Dynkin's formula~\cite[pp. 10]{nla.cat-vn712853} to (\ref{ditdis_tochu_general}), we have
\begin{align}
&E[V(x,\partial_{\mu} y,t)]-E[V(x(0),\partial_{\mu} y(0),0)]  \\
&\leq\! E\!\!\left[\int_0^t\!\!\Bigl(\!-2\alpha_1\|x(\tau)\!-\!x_d(\tau)\|^2\!\!+\!\underline{m}C\! +\!\frac{\overline{m}}{\varepsilon_1}\|d(x(\tau),\tau)\|^2\!\Bigr)d\tau\!\right]\nonumber
\end{align}
Using $E[V(x,\partial_{\mu} y,t)] > 0$ and $V(x(0),\partial_{\mu} y(0),0) = \|x(0)-x_d(0)\|_{M(0)}^2$ yields the desired inequality (\ref{ds_robust}).
\end{IEEEproof}
\begin{remark}
Corollary~\ref{robust_coro} implies that the CV-STEM control law yields  finite-gain $\mathcal{L}_2$ stability and input-to-state stability (ISS) in a mean squared sense (see Lemma~4 in~\cite{CHUNG20131148}). However, unlike the deterministic case, where $dV^p/dt = pV^{p-1}dV/dt$ can be used to prove the finite-gain $\mathcal{L}_p$ stability for $p \in [1,\infty)$, we have $\mathscr{L}V \neq pV^{p-1}\mathscr{L}V$. Directly computing $\mathscr{L}V^p$ using (\ref{l}) gives us the stability property of the proposed controller for general $p$ but it is left as future work due to space limitations.
\end{remark}
\subsection{ConVex optimization-based Stochastic steady-state Tracking Error Minimization (CV-STEM) Control}
\label{optimal_controller_design}
We formulate a convex optimization problem to find the optimal contraction metric $M(x,t)$, which greedily minimizes an upper bound of the steady-state mean squared distance in (\ref{stochasticmse}) of Theorem~\ref{thmcontracting}. This choice of $M(x,t)$ makes the stabilizing feedback control scheme (\ref{controller}) optimal in some sense.
\begin{assumption}
\label{as_ocd} From now on, we assume the following.
\begin{itemize}
    \item $\alpha$ and $\varepsilon$ are selected by a user. In particular, $\varepsilon$ can be chosen in a way that it minimizes the steady-state bound as explained in Remark \ref{epsilon_remark}.
    \item $\alpha_g$, which is defined below (\ref{alpha_cond}), is fixed; \ie, $\overline{m}_{x}$, $\overline{m}_{x^2}$, and $g_u$ are given.
    \item The upper bound of (\ref{stochasticmse}) as $t \to \infty$ is minimized instead of (\ref{stochasticmse}) itself.
    \item The objective value is minimized greedily at each step.
\end{itemize}
\end{assumption}
\subsubsection{Objective Function}
As a result of Theorem~\ref{thmcontracting}, we have
\begin{align}
\label{stochasticmse_infty}
&\lim_{t \to \infty}E\left[\|x_d-x\|^2\right] \leq \frac{C}{2\alpha} = \frac{g_u^2}{2\alpha}\left(\frac{\overline{m}}{\underline{m}}+c_1\frac{1}{\underline{m}}\right)
\end{align}
where $c_1 = \overline{m}_{x}/\varepsilon$. Since $\underline{m} = \inf_{x,t}\lambda_{\min}(M(x,t))$ and $\overline{m} = \sup_{x,t}\lambda_{\max}(M(x,t))$ depend on the future values of $M(x,t)$, the problem of directly minimizing (\ref{stochasticmse_infty}) becomes an infinite horizon problem. Instead of solving it, we greedily minimize the current steady-state upper bound (\ref{stochasticmse_infty}) to find an optimal $M(x,t)$ at the current time step as stated in Assumption \ref{as_ocd}. Namely, we drop $\inf$ and $\sup$ in the objective function (\ref{stochasticmse_infty}). The following lemma is critical in deriving the CV-STEM control framework.
\begin{lemma}
\label{new_objective_lemma}
The greedy objective function, \ie{}, the value inside the bracket of (\ref{stochasticmse_infty}) without $\inf$ and $\sup$, is upper bounded as follows:
\begin{align}
\label{new_objective}
\frac{\lambda_{\max}(M)}{\lambda_{\min}(M)}+\frac{c_1}{\lambda_{\min}(M)} \leq \kappa(W)+c_1{\kappa(W)^2}{\lambda_{\min}(W)}
\end{align}
where $W(x,t) = M(x,t)^{-1}$ and $\kappa(\cdot)$ is the condition number.
\end{lemma}
\begin{IEEEproof}
Rewriting the left-hand side of (\ref{new_objective}) using $\kappa$ gives
\begin{align}
\label{kappa_tochu}
\frac{\lambda_{\max}(M)}{\lambda_{\min}(M)}+\frac{c_1}{\lambda_{\min}(M)}\leq\kappa(M)+c_1\frac{\kappa(M)^2}{\lambda_{\max}(M)}
\end{align}
where $1 \leq \kappa(M) \leq \kappa(M)^2,\forall M$ by definition of $\kappa$ is used to upper-bound the term $c_1{\kappa(M)}/{\lambda_{\max}(M)}$. Substituting $\kappa(M) = \kappa(W)$ and $\lambda_{\max}(M) = 1/\lambda_{\min}(W)$ into (\ref{kappa_tochu}) completes the proof.
\end{IEEEproof}
\begin{remark}
We saw that the steady-state tracking error as a result of discrete-time stochastic contraction analysis in Theorem~\ref{disc_sic} is also a function of the condition number of the metric $M_k(x_k,t_k)$. This fact with the result of Theorem~\ref{disc_cont_thm} justifies the continuous-time control design to minimize the objective function written by the condition number of the metric $M(x,t)$, although the optimization-based controller has to be implemented in a discrete way in practical applications.
\end{remark}
\subsubsection{Convex Constraints}
Let us introduce additional variables $\chi$ and $\nu$ defined as 
\begin{align}
\label{lambda_con}
I \preceq \tilde{W} \preceq \chi I
\end{align}
where $\tilde{W}=\nu W$ and $\nu>0$.
\begin{lemma}
\label{equiv_constraints}
Suppose that the coefficients of the SDC parameterizations $\varrho$ are fixed. Given a positive constant $\nu$, the SDRI constraint (\ref{riccati}) is equivalent to the following convex constraint:
\begin{align}
\label{riccati_lmi_con}
-\dot{\tilde{W}}+A\tilde{W}+\tilde{W}A^{\top}+\tilde{\gamma} I-\nu BR^{-1}B^{\top} \preceq 0
\end{align}
where $\tilde{\gamma}=\nu \gamma$.
Similarly, the constraint (\ref{alpha_cond}) is equivalent to the following LMI constraint:
\begin{align}
\label{alpha_cond_convex_2}
\begin{bmatrix}
\tilde{\gamma} I+\nu BR^{-1}B^{\top}-\tilde{W}\phi^{\top}-\phi \tilde{W}-2\alpha \tilde{W}& \tilde{W} \\
\tilde{W} & \frac{\nu}{2\alpha_g}I \end{bmatrix} \succeq 0.
\end{align}
\end{lemma}
\begin{IEEEproof}
Since $\nu > 0$ and $W(x,t) \succ 0$, multiplying (\ref{riccati}) and (\ref{alpha_cond}) by $\nu$ and then by $W(x,t)$ from both sides preserves matrix definiteness. Also, the resultant inequalities are equivalent to the original ones~\cite[pp. 114]{lmi}. For the SDRI constraint (\ref{riccati}), these operations yield the desired inequality (\ref{riccati_lmi_con}).
For the constraint (\ref{alpha_cond}), these operations give us that
\begin{align}
\label{cont_con_tochu}
&\tilde{\gamma} I+\nu BR^{-1}B^{\top}-\tilde{W}\phi^{\top}-\phi \tilde{W}- \frac{2\alpha_g}{\nu}\tilde{W}^2 \succeq 2\alpha \tilde{W}.
\end{align}
Applying Schur's complement lemma~\cite[pp. 7]{lmi} to (\ref{cont_con_tochu}) results in the desired LMI constraint (\ref{alpha_cond_convex_2}).
\end{IEEEproof}

\subsubsection{Convex Optimization Formulation}
We are now ready to state our main result on convex optimization-based sampling of optimal contraction metrics.
\begin{theorem}
\label{convex_equiv}
Suppose $\alpha$, $g_u$, and $c_1$ in (\ref{stochasticmse_infty}) are given. Then the non-convex optimization problem of greedily minimizing a steady-state upper bound of $E[\|x-x_d\|^2]$ in Theorem~\ref{thmcontracting} is defined as follows:
\begin{align}
\label{problem1}
&\mathcal{J}_{nl}^* = \min_{\gamma>0,W\succ0,M\succ0} \kappa(W)+c_1{\kappa(W)^2}{\lambda_{\min}(W)} \\
&\text{\st{} }\text{(\ref{riccati}), (\ref{alpha_cond}), and $M(x,t) = W(x,t)^{-1}$}. \nonumber
\end{align}
Further, the following convex optimization problem
\begin{align}
\label{convex_opt_gen}
&\mathcal{J}_{cv}^* = \min_{\substack{\tilde{\gamma}>0,\nu>0,\tau\in\mathbb{R}\\\chi\in \mathbb{R},\tilde{W}\succ0}}\tau \\
&\text{\st{} }\text{(\ref{lambda_con}), (\ref{riccati_lmi_con}), (\ref{alpha_cond_convex_2}), and }
\begin{bmatrix}
\tau-\chi & \chi \\
\chi & \frac{\nu}{c_1}
\end{bmatrix} \succeq 0. \nonumber
\end{align}
is equivalent to the non-convex counterpart (\ref{problem1}), \ie{}, $\mathcal{J}_{nl}^*=\mathcal{J}_{cv}^*$.
\end{theorem}
\begin{IEEEproof}
The first part (\ref{problem1}) follows from Lemma~\ref{new_objective_lemma}, which derives an upper bound of the steady-state mean squared distance (\ref{stochasticmse}) under the conditions (\ref{riccati}) and (\ref{alpha_cond}).
For the second part, consider the following two optimization problems:
\begin{align}
\label{problem2}
&\mathcal{J}_{n2c}^* = \min_{\substack{\tilde{\gamma}>0,\nu>0\\\chi\in \mathbb{R},\tilde{W}\succ0}}\chi+c_1\frac{\chi^2}{\nu} \text{~~\st~(\ref{lambda_con}), (\ref{riccati_lmi_con}), and (\ref{alpha_cond_convex_2})}
\end{align}
and
\begin{align}
\label{problem3}
&\hat{\mathcal{J}}_{n2c}^* = \min_{\substack{\tilde{\gamma}>0,\nu>0\\\chi\in \mathbb{R},\tilde{W}\succ0}}\chi+c_1\frac{\chi^2}{\nu} \\
&\text{\st~(\ref{riccati_lmi_con}), (\ref{alpha_cond_convex_2}), } \lambda_{\min}(\tilde{W}) = 1,\text{and }\lambda_{\max}(\tilde{W}) = \chi. \nonumber
\end{align}
The rest of the proof is outlined as follows: we first prove $\mathcal{J}_{nl}^* = \mathcal{J}_{n2c}^*$ by showing a) $\mathcal{J}_{nl}^* = \hat{\mathcal{J}}_{n2c}^* \geq \mathcal{J}_{n2c}^*$ and b) $\mathcal{J}_{nl}^* \leq \mathcal{J}_{n2c}^*$, and then prove c) $\mathcal{J}_{n2c}^*=\mathcal{J}_{cv}^*$ to obtain the desired relation $\mathcal{J}_{nl}^*=\mathcal{J}_{cv}^*$.

a) $\mathcal{J}_{nl}^* = \hat{\mathcal{J}}_{n2c}^* \geq \mathcal{J}_{n2c}^*$: Let us denote the feasible set of (\ref{problem2}) as $\mathcal{S}_{n2c}$ and that of (\ref{problem3}) as $\mathcal{\hat{S}}_{n2c}$. Due to the constraint (\ref{lambda_con}), which can be rewritten as $\lambda_{\min}(\tilde{W}) \geq 1$ and $\lambda_{\max}(\tilde{W}) \leq \chi$, we have $\mathcal{\hat{S}}_{n2c} \subseteq \mathcal{S}_{n2c}$. This indicates that $\mathcal{\hat{J}}_{n2c}^* \geq \mathcal{{J}}_{n2c}^*$ as (\ref{problem2}) and (\ref{problem3}) use the same objective function. Also, using $\nu = 1/\lambda_{\min}(W)$ and $\chi = \lambda_{\max}(\tilde{W}) = \kappa(W)$, $\forall \nu,\chi \in \mathcal{\hat{S}}_{n2c}$ by definition, $\mathcal{\hat{J}}_{n2c}^*$ can be expressed as
\begin{align}
\label{problem4}
&\hat{\mathcal{J}}_{n2c}^* = \min_{\substack{\tilde{\gamma}>0,\nu>0,\tilde{W}\succ0}}\kappa(W)+c_1{\kappa(W)^2}{\lambda_{\min}(W)} \\
&\text{\st~(\ref{riccati_lmi_con}) and (\ref{alpha_cond_convex_2})}. \nonumber
\end{align}
Since (\ref{riccati_lmi_con}) and (\ref{alpha_cond_convex_2}) are equivalent to (\ref{riccati}) and (\ref{alpha_cond}), respectively, as proved in Lemma~\ref{equiv_constraints}, (\ref{problem1}) and (\ref{problem4}) imply that $\mathcal{J}_{nl}^* = \hat{\mathcal{J}}_{n2c}^*$. Thus, we have $\mathcal{J}_{nl}^* = \hat{\mathcal{J}}_{n2c}^* \geq \mathcal{J}_{n2c}^*$ as desired.

b) $\mathcal{J}_{nl}^* \leq \mathcal{J}_{n2c}^*$: For $\tilde{W} \in \mathcal{S}_{n2c}$, we have
\begin{align}
\label{eq61}
&\kappa({W})+{c_1}{\kappa({W})^2}{\lambda_{\min}({W})} = \frac{\lambda_{\max}(\tilde{W})}{\lambda_{\min}(\tilde{W})}+{c_1}\frac{(\lambda_{\max}(\tilde{W}))^2}{\nu\lambda_{\min}(\tilde{W})} \nonumber \\
&\leq  {\lambda_{\max}(\tilde{W})}+{c_1}\frac{(\lambda_{\max}(\tilde{W}))^2}{\nu} \leq {\chi}+{c_1}\frac{\chi^2}{\nu}.
\end{align}
where $\kappa(W) = \kappa(\tilde{W})$ and $\lambda_{\min}(W) = \lambda_{\min}(\tilde{W})/\nu$ are used for the first equality, and (\ref{lambda_con}) expressed as $\lambda_{\min}(\tilde{W}) \geq 1$ and $\lambda_{\max}(\tilde{W}) \leq \chi$ is used for the second and third inequalities, respectively. 
Since (\ref{eq61}) holds for any decision variable in $\mathcal{S}_{n2c}$, we have $\mathcal{{J}}_{nl}^* \leq \mathcal{J}_{n2c}^*$ by (\ref{problem1}) and (\ref{problem2}).

c) $\mathcal{J}_{n2c}^*=\mathcal{J}_{cv}^*$: The epigraph form~\cite[pp. 134]{citeulike:163662} of (\ref{problem2}) is given as
\begin{align}
\label{problem5}
&\mathcal{J}_{n2c}^* = \min_{\substack{\tilde{\gamma}>0,\nu>0,\tau\in\mathbb{R}\\\chi\in \mathbb{R},\tilde{W}\succ0}}\tau \\
&\text{\st{} }\text{(\ref{lambda_con}), (\ref{riccati_lmi_con}), (\ref{alpha_cond_convex_2}), and }\tau \geq \chi+c_1 \frac{\chi^2}{\nu} \nonumber
\end{align}
Applying Schur's complement lemma~\cite[pp. 7]{lmi} to the last constraint of (\ref{problem5}) results in $\mathcal{J}_{n2c}^*=\mathcal{J}_{cv}^*$.
\end{IEEEproof}
\begin{remark}
Although (\ref{convex_opt_gen}) is convex, it is infinite dimensional due to $\dot{\tilde{W}}$. We could address this issue by computing $\dot{\tilde{W}}$ along the trajectory or by approximating the contraction metric as a linear combination of given basis functions~\cite{AYLWARD20082163}. These techniques will be briefly discussed in Sec.~\ref{variations}.
\end{remark}

The coefficients of the SDC parameterizations $\varrho$ can also be treated as a decision variable as can be seen in the following proposition.
\begin{proposition}
\label{relaxed_convex}
Introducing new variables $\tilde{W}_{\varrho_i}\succ 0$ and $\tilde{\varrho_i}=\nu{\varrho_i}$ where $\tilde{W}_{\varrho_i} = \varrho_i \tilde{W}$, the bilinear matrix inequalities (\ref{riccati_lmi_con}) and (\ref{alpha_cond_convex_2}) in terms of $\tilde{W}$ and $\varrho$ with $\nu > 0$ can be relaxed as follows:
\begin{align}
\label{riccati_lmi_con_rho}
-\dot{\tilde{W}}+{\displaystyle \sum_{i=1}^{s_1}}A_i\tilde{W}_{\varrho_i}+{\displaystyle \sum_{i=1}^{s_1}}\tilde{W}_{\varrho_i}A_i^{\top}+\tilde{\gamma} I-\nu BR^{-1}B^{\top} \preceq 0
\end{align}
and
\begin{align}
\label{alpha_cond_convex_2_rho}
\begin{bmatrix}
\tilde{\gamma} I+\nu BR^{-1}B^{\top}-\Phi-2\alpha \tilde{W}& \tilde{W} \\
\tilde{W} & \frac{\nu}{2\alpha_g}I \end{bmatrix} \succeq 0.
\end{align}
where $\Phi$ is given by
\begin{align}
\Phi =& {\displaystyle \sum_{i=1}^{s_1}}\tilde{W}_{\varrho_i}\frac{\partial (\Delta A_ix_d) }{\partial q}^{\top}+{\displaystyle \sum_{i=1}^{s_1}}\frac{\partial (\Delta A_ix_d) }{\partial q} \tilde{W}_{\varrho_i} \nonumber \\
&+\tilde{W}\frac{\partial (\Delta Bu_d)}{\partial q}^{\top}+\frac{\partial (\Delta Bu_d)}{\partial q}\tilde{W} \nonumber
\end{align}
with $\Delta A(\varrho,x,t) = \sum_{i=1}^{s_1}\varrho_i \Delta A_i(x,t) = \sum_{i=1}^{s_1}\varrho_i (A_i(x,t)-A_i(x_d,t))$.
We also need some additional relaxed constraints to ensure controllability and $\tilde{W}_{\varrho_i} = \varrho_i \tilde{W}$, \ie,
\begin{align}
\label{last_constraint_con}
&\tilde{W},\tilde{W}_{\varrho_i} \succ 0,~\sum_{i=1}^{s_1}\tilde{W}_{\varrho_i}=\tilde{W}, 
~{\rm sym}
\begin{bmatrix}
\nu I & \tilde{W} \\
\tilde{\varrho}_i I & \tilde{W}_{\varrho_i} \end{bmatrix} \succeq 0, \\
&\sum_{i=1}^{s_1}\tilde{\varrho}_i=\nu,~\tilde{\varrho}_i \in [0,\nu],~cc_k(\tilde{\varrho},x)\leq0,~\forall i,~\forall k=1,\cdots, n_c \nonumber
\end{align}
where $cc_k(\tilde{\varrho},x)\leq0,~\forall k=1,\cdots, n_c$ denotes convex constraints to maintain the controllability of the pair $(A,B)$.
\end{proposition}
\begin{IEEEproof}
The first two inequalities (\ref{riccati_lmi_con_rho}) and (\ref{alpha_cond_convex_2_rho}) follow from the desired equality $\tilde{W}_{\varrho_i} = \varrho_i\tilde{W}$ and $A(\varrho,x,t)=\sum_{i=1}^{s_1}\varrho_iA_i(x,t)$. See~\cite{observer} for the derivation of (\ref{last_constraint_con}).
\end{IEEEproof}
\subsubsection{Summary of CV-STEM Control Design}
\label{optcon}
The CV-STEM control of a class of It\^{o} stochastic nonlinear systems is designed as (\ref{controller}), where the optimal contraction metric $M(x) = \nu \tilde{W}(x)^{-1}$ is selected by the convex optimization problem (\ref{convex_opt_gen}) in Theorem~\ref{convex_equiv}.
The coefficients of SDC parameterizations $\varrho$ can also be used to preserve controllability by considering the relaxed problem with the constraints (\ref{riccati_lmi_con_rho}), (\ref{alpha_cond_convex_2_rho}), and (\ref{last_constraint_con}) in Proposition~\ref{relaxed_convex}, where the decision variables are $\tilde{\gamma} > 0$, $\nu\in\mathbb{R}$, $\tau\in\mathbb{R}$, $\chi\in\mathbb{R}$, $\tilde{W} \succ 0$, $\tilde{W}_{\varrho_i}\succ 0$, and $\tilde{\varrho_i}$.

The CV-STEM control design provides a convex optimization-based methodology for computing the contraction metric that greedily minimizes an upper bound of the steady-state mean squared tracking error (\ref{stochasticmse}) in Theorem~\ref{thmcontracting}. As proved in Corollary~\ref{robust_coro}, it is also robust against external disturbances and has the $\mathcal{L}_2$ norm bound on the tracking error. In practice, (\ref{convex_opt_gen}) of Theorem~\ref{convex_equiv} can be implemented using computationally-efficient numerical techniques such as the polynomial-time interior point method for convex programming~\cite{lmi, citeulike:163662, Ben-Tal:2001:LMC:502969,5447065} and the SDRI solvers~\cite{1102178,1098829,1100565,1101323,Anderson:1990:OCL:79089}. Although the control parameters are supposed to be updated by (\ref{convex_opt_gen}) at each time instant due to the state-and time-dependent constraints, its sampling period can be relaxed to larger values to allow online implementation of the CV-STEM as shall be seen in  Sec.~\ref{simulation}. Further, the controllability constraint can be incorporated into this framework~\cite{observer} as in Proposition~\ref{relaxed_convex}, utilizing non-unique choices of SDC parametrizations.
\section{Main Result 3: CV-STEM Control Design for Lagrangian Systems}
\label{lagrangian_systems}
We consider stochastic Lagrangian systems equipped with an exponentially-stabilizing tracking controller~\cite{Slotine:1228283}. We propose a robust optimization-based controller that can handle stochastic disturbances and guarantee exponential boundedness of the mean squared tracking error of system trajectories.
\subsection{Problem Formulation and Feedback Control Design}
Let us consider the following Lagrangian system with a stochastic disturbance:
\begin{align}
\label{sto_lag_system}
\mathcal{H}(q)d \dot{q}+(\mathcal{C}(q,\dot{q})\dot{q}+\mathcal{G}(q))dt=\mathcal{B}(q,\dot{q})u dt+\Gamma(x,t)d\mathscr{W}
\end{align}
where $q:\mathbb{R}_{\geq0} \to \mathbb{R}^{n}$, $u:\mathbb{R}_{\geq0} \to \mathbb{R}^{m}$, $\mathcal{H}:\mathbb{R}^n \to \mathbb{R}^{n\times n}$, $\mathcal{C}:\mathbb{R}^n \times \mathbb{R}^n \to \mathbb{R}^{n\times n}$, $\mathcal{G}:\mathbb{R}^n \to \mathbb{R}^{n}$, $\mathcal{B}:\mathbb{R}^n \times \mathbb{R}^n \to \mathbb{R}^{n\times m}$, and $\Gamma:\mathbb{R}^n \times \mathbb{R}_{\geq0} \to \mathbb{R}^{n\times d}$ with the same assumptions on the existence and uniqueness of the solution stated in Sec.~\ref{preliminaries}. We note that the matrix $\mathcal{C}(q,\dot{q})$ is selected to make $\dot{\mathcal{H}}-2\mathcal{C}$ skew-symmetric, so we have a useful property \st{} $z^{\top}(\dot{\mathcal{H}}-2\mathcal{C})z=0,~\forall z\in\mathbb{R}^n$. A feedback controller $u$ for this system is designed as a combination of an exponentially stabilizing nominal controller $u_n$ and a stochastic controller $u_s$:
\begin{align}
\label{lag_con}
u &= u_{n}+u_{s} \\
u_{n} &= \mathcal{B}(q,\dot{q})^{+}(\mathcal{H}(q)\ddot{q}_r+\mathcal{C}(q,\dot{q})\dot{q}_r+\mathcal{G}(q)-\mathcal{K}(t)(\dot{q}-\dot{q}_r)) \nonumber \\
u_s &= -K_s(x)s = -R(x)^{-1}B(x)^{\top}M(x)s \nonumber
\end{align}
where $\dot{q}_r = \dot{q}_d-\Lambda (q-q_d)$, $s=\dot{q}-\dot{q}_r$, $x = [q^{\top},\dot{q}^{\top}]^{\top}$, and
\begin{align}
&A(x) = -\mathcal{H}(q)^{-1}(\mathcal{C}(q,\dot{q})+\mathcal{K}(t)) \nonumber \\
&B(x) = \mathcal{H}(q)^{-1}\mathcal{B}(q,\dot{q})\nonumber \\
\label{sdri}
&\dot{M}+MA+A^{\top}M-MBR^{-1}B^{\top}M+\gamma M^2 \preceq 0.
\end{align}
with $M \succ 0$ and $\gamma > 0$. $R(x) \succ 0$ is a weight matrix on the input $u_s$.
When $\mathcal{B}\mathcal{B}^{+} = I$, applying (\ref{lag_con}) to (\ref{sto_lag_system}) yields the following closed loop system:
\begin{align}
&\mathcal{H}(q)d{s}+(\mathcal{C}(q,\dot{q})+\mathcal{K}(t))sdt \nonumber \\
&=-\mathcal{B}(q,\dot{q})K_s(x)sdt+\Gamma(x,t)d\mathscr{W}.
\end{align}
\begin{remark}
In the proceeding stability proof in Theorem~\ref{thm_lag_sto}, the metric $M$ in (\ref{sdri}) is a contraction metric to handle stochasticity in the Lagrangian system, while the inertia matrix $\mathcal{H}$ is for guaranteeing deterministic contraction.
\end{remark}
\begin{lemma}
\label{lemma_det_robustness}
Suppose that the deterministic system is perturbed as follows:
\begin{align}
\mathcal{H}(q)\dot{s}+(\mathcal{C}(q,\dot{q})+\mathcal{K}(t))s=\mathcal{B}(q,\dot{q})(u_s+d).
\end{align}
If there exists a positive definite solution $M(x)$ to (\ref{sdri})
with $R(x) = S(x)^2 \succ 0$ and $S(x) \succ 0$, then the system with an input $\mu = S(x)d$ and an output $y = \sqrt{\gamma}M(x)s$ is finite-gain $\mathcal{L}_2$ stable and its $\mathcal{L}_2$ gain is less than or equal to 1.
\end{lemma}
\begin{IEEEproof}
Following the same proof as in Appendix~\ref{proof_det_robustness} with the Lyapunov function $V_M = s^{\top}Ms$, we have $\dot{V}_M \leq -\|y\|^2+\|\mu\|^2$ due to (\ref{sdri}). This relation along with the comparison lemma~\cite[pp. 211]{Khalil:1173048} gives us the desired result.
\end{IEEEproof}
\begin{remark}
Since the system with the output $y = \sqrt{\gamma}M(x)s$ and input $\mu=S(x)d$ is clearly zero-state observable~\cite{256331}, it is exponentially stable when $d = 0$.
\end{remark}
\subsection{Incremental Stability Analysis}
Let us define a virtual system of (\ref{sto_lag_system}) as follows:
\begin{align}
\label{vsystem_stochastic_lag}
&\mathcal{H}(q)d{y}+(\mathcal{C}(q,\dot{q})+\mathcal{K}(t))y(\mu,t)dt \nonumber \\ 
&= -\mathcal{B}(q,\dot{q})K_s(x)y(\mu,t)dt+\Gamma_y(y(\mu,t),t)d\mathscr{W}
\end{align}
where $\mu \in [0,1]$ is introduced to parameterize the trajectories $y=s$ and $y=0$, \ie, $y(\mu=0,t) = s$, $y(\mu=1,t) = 0$, $\Gamma_y(y(0,t),t) = \Gamma(x,t)$, and $\Gamma_y(y(1,t),t) = 0_{n \times d}$. Note that (\ref{vsystem_stochastic_lag}) has $y=s$ and $y=0$ as particular solutions as a result of this parameterization.  The following theorem analyzes a stochastic contraction property of the Lagrangian system (\ref{sto_lag_system}) under the feedback control (\ref{lag_con}) similarly to Theorem~\ref{thmcontracting}.
\begin{theorem}
\label{thm_lag_sto}
Suppose there exist $\overline{\ell}_{x}$, $\overline{\ell}_{x^2}$, and $g_B$ \st{} $\|\mathcal{H}(q)^{-1}\Gamma(x,t)\|_F\leq g_B$, $\|\partial ((\mathcal{H}(q)+\sigma M(x))_{ij})/\partial x\|\leq\overline{\ell}_{x}$, and $\left\|{\partial^2((\mathcal{H}(q)+\sigma M(x))_{ij})}/{\partial x^2}\right\|\leq\overline{\ell}_{x^2}$, $\forall x$, where $\overline{\ell}_{x}$, $\overline{\ell}_{x^2}$, and $g_B$ are bounded.
Suppose also that there exist $\alpha_{\ell}>0$ and $\sigma>0$ \st{}
\begin{align}
\label{contcon_sto}
&\mathcal{B}(q,\dot{q})R(x)^{-1}B(x)^{\top}M(x)+M(x)B(x)R(x)^{-1}\mathcal{B}(q,\dot{q})^{\top} \nonumber \\
&+\sigma(\gamma M(x)^2+M(x)B(x)R(x)^{-1}B(x)^{\top}M(x))-2\alpha_{\gamma} I \nonumber \\ 
&\succeq 2\alpha_{\ell} (\mathcal{H}(q)+\sigma M(x))
\end{align}
where $2\alpha_{\gamma} = g_B^2\left(\overline{\ell}_{x}\varepsilon_{\ell}+\overline{\ell}_{x^2}/2\right)$ with an arbitrary positive constant $\varepsilon_{\ell}$.
If there exists a positive definite solution $M(x,t)$ to the inequalities (\ref{sdri}) and (\ref{contcon_sto}), then the mean-squared distance of the composite state $s$ is bounded as follows:
\begin{align}
\label{sbound}
E[\|s\|^2] \leq \frac{E[V(x(0),\partial_{\mu} y(0),0)]e^{-2\overline{\alpha}t}+\frac{C_{\ell}}{2 \overline{\alpha}}}{\inf_{t \geq 0}\lambda_{\min}(\mathcal{H}(q)+\sigma M(x))}
\end{align}
where $V(x,\partial_{\mu} y)$ is given by
\begin{align}
\label{lag_lyap}
V(x,\partial_{\mu} y) = \int_0^1\frac{\partial y}{\partial \mu}^{\top} (\mathcal{H}(q)+\sigma M(x)) \frac{\partial y}{\partial \mu}d \mu
\end{align}
with $C_{\ell} = g_B^2\sup_{t\geq0}(\lambda_{\max}(\mathcal{H}(q)+\sigma M(x)))+{\overline{\ell}_{x}g_{B}^2}/{\varepsilon_{\ell}}$, $\overline{\alpha} = \alpha_{\ell}+\underline{k}/\sup_t\lambda_{\max}(\mathcal{H}(q)+\sigma M(x))$, and $\underline{k}I \prec \mathcal{K}(t), \forall t$.
\end{theorem}
\begin{IEEEproof}
Following the same proof given in Theorem~\ref{thmcontracting}, the condition (\ref{sdri}) gives us that
\begin{align}
\label{test}
&\mathscr{L}V
\leq-\sigma\int_0^1\frac{\partial y}{\partial \mu}^{\top}(\gamma M^2+MBR^{-1}B^{\top}M) \frac{\partial y}{\partial \mu} d\mu \\
&-2\int_0^1\frac{\partial y}{\partial \mu}^{\top} (\mathcal{K}+\mathcal{B}K_s)\frac{\partial y}{\partial \mu} d\mu+2\alpha_{\gamma}\int_0^1\left\|\frac{\partial y}{\partial \mu}\right\|^2d\mu+C_{\ell} \nonumber
\end{align}
where the skew-symmetric property of $\dot{\mathcal{H}}-2\mathcal{C}$ is used to obtain the above inequality.
Using (\ref{contcon_sto}), we have
\begin{align}
\mathscr{L}V \leq& -2\alpha_{\ell} V-2\int_0^1\frac{\partial y}{\partial \mu}^{\top} \mathcal{K}\frac{\partial y}{\partial \mu} d\mu +C_{\ell} \nonumber \\
\leq& -2\overline{\alpha} V+C_{\ell}.
\end{align}
Thus, applying Theorem~\ref{sic} yields the desired result (\ref{sbound}).
\end{IEEEproof}

\subsection{Robustness against Stochastic and Deterministic Disturbances}
Analogously to Lemma~\ref{lemma_det_robustness}, consider the following Lagrangian system with deterministic and stochastic disturbances:
\begin{align}
&\mathcal{H}(q)d \dot{q}+(\mathcal{C}(q,\dot{q})\dot{q}+\mathcal{G}(q))dt \nonumber \\
&= \mathcal{B}(q,\dot{q})q dt+d(x,t)dt+\Gamma(q,\dot{q})d\mathscr{W}.
\end{align}
Again, an important example of these systems is a parametric uncertain system.
\begin{corollary}
Let $\mathcal{H}_0 = \mathcal{H}(0)+\sigma M(0)$. The controller (\ref{lag_con}) with the constraints (\ref{sdri}) and (\ref{contcon_sto}) is robust against the external disturbances and satisfies the following $\mathcal{L}_2$ norm bound on the tracking error:
\begin{align}
\label{ds_robust_lagrange}
&E[\|(s)_{\tau}\|^2_{\mathcal{L}_2}] \leq \frac{E[\|s(0)\|_{\mathcal{H}_0}^2]+\frac{\overline{\ell}}{\underline{\varpi}\varepsilon_2}E[\|(d)_{\tau}\|^2_{\mathcal{L}_2}]+C_{\ell}\tau}{2\alpha_2}
\end{align}
where $\underline{\ell}I \preceq \mathcal{H}(q)+\sigma M(x) \preceq \overline{\ell}I$, $\underline{\varpi}I \preceq \mathcal{H}(q)$, $\forall x$, and $\alpha_2 = \overline{\alpha}\underline{\ell}-\varepsilon_2{\overline{\ell}}/(2\underline{\varpi})$ with $\varepsilon_2>0$ that guarantees $\alpha_2 > 0$.
\end{corollary}
\begin{IEEEproof}
Following the same proof as in Corollary~\ref{robust_coro}, we have $\mathcal{L}V\leq -2\alpha_2\int_{0}^1\|\partial y/\partial \mu\|^2d\mu+\overline{\ell}/(\underline{\varpi}\varepsilon_2)\|d(x,t)\|^2+C_{\ell}$, where $y$ is the virtual state and $V$ is given in (\ref{lag_lyap}). The rest follows from Dynkin's formula~\cite[pp. 10]{nla.cat-vn712853}.
\end{IEEEproof}
\subsection{Convex Optimization Formulation}
As a result of Theorem~\ref{thm_lag_sto}, we have
\begin{align}
\label{sboundinf}
\lim_{t \to \infty}E[\|s\|^2] \leq \frac{g_B^2\sup_{t\geq0}(\lambda_{\max}(\mathcal{H}+\sigma M))+\frac{\overline{\ell}_{x}g_{B}^2}{\varepsilon_{\ell}}}{2 \overline{\alpha}\inf_t\lambda_{\min}(\mathcal{H}+\sigma M)}.
\end{align}
We propose one way to formulate a convex optimization problem to find the optimal contraction metric which minimizes an upper bound of the right-hand side of (\ref{sboundinf}) under the following conditions.
\begin{assumption}
\label{lag_as}
In addition to the conditions given in Assumption \ref{as_ocd}, we assume that $\sigma = 1$, which is possible as we can optimally select the value of $\gamma$.
\end{assumption}
\subsubsection{Objective Function}
Under Assumption~\ref{lag_as}, we have the following lemma on the greedy objective function as in Lemma~\ref{new_objective_lemma} of Sec.~\ref{optimal_controller_design}.
\begin{lemma}
\label{ob_upper_bound_lemma}
The greedy objective function, \ie, (\ref{sboundinf}) without $\sup$, $\inf$, and with $g_B$ and $\overline{\alpha}$ given, is bounded as follows:
\begin{align}
\label{ob_upper_bound}
\frac{\lambda_{\max}(\mathcal{H}+M)+\frac{\overline{\ell}_{x}}{\varepsilon_{\ell}}}{\lambda_{\min}(\mathcal{H}+M)} \leq \kappa(W)+c_2{\kappa(W)^2}{\lambda_{\min}(W)}
\end{align}
where $W(x) = M(x)^{-1}$ and $c_2 = \lambda_{\max}(\mathcal{H})+\overline{\ell}_{x}/\varepsilon_{\ell}$.
\end{lemma}
\begin{IEEEproof}
Using the relations $\lambda_{\max}(\mathcal{H}+M) \leq \lambda_{\max}(\mathcal{H})+\lambda_{\max}(M)$ and  $\lambda_{\min}(\mathcal{H}+M) \geq \lambda_{\min}(\mathcal{H})+\lambda_{\min}(M)\geq \lambda_{\min}(M)$~\cite[pp. 242]{10.5555/2422911}, we have
\begin{align}
\label{lag_tochu}
\frac{\lambda_{\max}(\mathcal{H}+M)+\frac{\overline{\ell}_{x}}{\varepsilon_{\ell}}}{\lambda_{\min}(\mathcal{H}+M)} \leq \frac{\lambda_{\max}(M)}{\lambda_{\min}(M)}+\frac{c_2}{\lambda_{\min}(M)}
\end{align}
Applying Lemma~\ref{new_objective_lemma} to (\ref{lag_tochu}) completes the proof.
\end{IEEEproof}
\subsubsection{Equivalent Convex Optimization Problem}
Let us introduce $\nu>0$, $\chi,\tau\in\mathbb{R}$, and $\tilde{W}=\nu W\succ 0$ constrained as
\begin{align}
\label{convex1}
I \preceq \tilde{W} \preceq \chi I,~~\begin{bmatrix}
\tau-\chi & \chi \\
\chi & \frac{\nu}{c_2}
\end{bmatrix} \succeq 0.
\end{align}
Analogously to Theorem~\ref{convex_equiv}, we have the following results.
\begin{theorem}
\label{lag_CVSTEM}
Suppose $\overline{\alpha}$, $g_B$, and $c_2$ are given. Then the non-convex optimization problem of greedily minimizing an upper bound of (\ref{sboundinf}) due to Theorem~\ref{thm_lag_sto} is defined as follows:
\begin{align}
\label{problemNL}
&\mathcal{J}_{nl\ell}^* = \min_{\gamma>0,W\succ0,M\succ0} \kappa(W)+c_2{\kappa(W)^2}{\lambda_{\min}(W)} \\
&\text{\st{} }\text{(\ref{sdri}), (\ref{contcon_sto}), and $M(x,t) = W(x,t)^{-1}$}. \nonumber
\end{align}
Further, the following convex optimization problem
\begin{align}
\label{op_convex}
&\mathcal{J}_{cv\ell}^* = \min_{\substack{\tilde{\gamma}>0,\nu>0,\tau \in \mathbb{R} \\\chi\in \mathbb{R},\tilde{W}\succ0}}\tau \\
\label{convex2}
&\text{\st{} }\dot{\tilde{W}}+A\tilde{W}+\tilde{W}A^{\top}-\nu BR^{-1}B^{\top}+\tilde{\gamma} I \preceq 0 \\
\label{convex3}
&\begin{bmatrix}
\tilde{\mathcal{H}}_{\ell} & \tilde{W} \\
\tilde{W} & \frac{\nu}{2}(\alpha_{\ell} \mathcal{H}+\alpha_{\gamma}I)^{-1}
\end{bmatrix}
\succeq 0\text{ and (\ref{convex1})}
\end{align}
where $\tilde{\gamma} = \nu\gamma$ and $\tilde{\mathcal{H}}_{\ell} = 2\sym(\tilde{W}\mathcal{B}R^{-1}B^{\top})+\tilde{\gamma} I+\nu BR^{-1}B^{\top}-2\alpha_{\ell} \tilde{W}$, is equivalent to (\ref{problemNL}), \ie{}, $\mathcal{J}_{nl\ell}^*=\mathcal{J}_{cv\ell}^*$.
\end{theorem}
\begin{IEEEproof}
The first part follows from Lemma~\ref{ob_upper_bound_lemma}. The constraints (\ref{sdri}) and (\ref{contcon_sto}) are equivalent to (\ref{convex2}) and the first constraint of (\ref{convex3}), respectively, as shown in Lemma~\ref{equiv_constraints}. The rest follows from the same proof as in Theorem~\ref{convex_equiv}.
\end{IEEEproof}
In summary, the CV-STEM control of stochastic Lagrangian systems is designed as (\ref{lag_con}), where the optimal contraction metric $M(x) = \nu \tilde{W}(x)^{-1}$ is selected by the convex optimization problem (\ref{op_convex}) in Theorem~\ref{lag_CVSTEM}.
\section{Main Result 4: CV-STEM with Input Constraints and Other Extensions}
\label{variations}
Several extensions of algorithms to compute the optimal contraction metric for the feedback control of It\^{o} stochastic nonlinear systems are discussed in this section.
\subsection{Input Constraints}
We propose two ways to incorporate input constraints into the convex optimization problem (\ref{convex_opt_gen}) of Theorem~\ref{convex_equiv} and (\ref{op_convex}) of Theorem~\ref{lag_CVSTEM} without losing their convexity.
\subsubsection{Input Constraints through the Feedback Gain}
Let us consider the case when the input constraint can be relaxed to $\|u(t)\| \leq u_{\max}$, where $u(t)$ is defined in (\ref{controller}) and $u_{\max}>0$ is given.
\begin{proposition}
\label{tot_input_con_prop}
A sufficient condition for the input constraint $\|u(t)\| \leq u_{\max}$, $\forall t\geq0$ with a given $u_{\max}(\geq\|u_d(t)\|)$ is expressed as follows:
\begin{align}
\label{tot_input_con}
\nu\|R^{-1}B^{\top}\|\|e(t)\| \leq (u_{\max}-\|u_d(t)\|)\lambda_{\min}(\tilde{W}),~\forall t,x
\end{align}
where $e(t) = x(t)-x_d(t)$ and the arguments $(x,t)$ are dropped for notational simplicity. Further, this is a convex constraint in terms of the decision variables of (\ref{convex_opt_gen}) in Theorem~\ref{convex_equiv}.
\end{proposition}
\begin{IEEEproof}
Using the relations $M=\nu\tilde{W}^{-1}$ and $\|\tilde{W}^{-1}\| \leq 1/\lambda_{\min}(\tilde{W})$ due to (\ref{lambda_con}), we have
\begin{align}
\|u\| =& \|-K(x-x_d)+u_d\| = \|\nu R^{-1}B^{\top}\tilde{W}^{-1} e\|+\|u_d\| \nonumber \\
\leq& \frac{\nu\|R^{-1}B^{\top}\|\|e\|}{\lambda_{\min}(\tilde{W})}+\|u_d\|.
\end{align}
Thus, a sufficient condition for $\|u(t)\| \leq u_{\max}$, $\forall t\geq0$ reduces to (\ref{tot_input_con}). Also, this is convex in terms of $\nu$ and $\tilde{W}$ as $u_{\max}-\|u_d\|\geq0$ by assumption and $\lambda_{\min}(\tilde{W})$ is a concave function~\cite[pp. 118]{citeulike:163662}.
\end{IEEEproof}
Proposition~\ref{tot_input_con_prop} allows us to implement $\|u(t)\| \leq u_{\max}$, $\forall t\geq0$ in (\ref{convex_opt_gen}) and (\ref{op_convex}) without losing their convexity.
\subsubsection{Input Constraints through CLFs}
Let us take (\ref{op_convex}) as an example. Although $u_s$ is given by $u_s = -K_s s$ in (\ref{lag_con}), this form of $u_s$ is not optimal in any sense. Instead, we find $u_s$ which minimizes its Euclidean norm, assuming $M(x,t)$ and $\gamma$ are obtained by solving (\ref{op_convex}). The following proposition allows us to optimally incorporate input constraints without dramatically changing the CV-STEM stability and optimality properties.
\begin{proposition}
\label{clfqp_incon_prop}
Consider the following convex optimization problem to minimize $\|u_s\|$ with an input constraint $u_s\in\mathcal{U}_s$, where $\mathcal{U}_s$ is a given convex set:
\begin{align}
\label{clfqp_incon}
&u_s^* = \text{arg}\min_{\substack{u_s \in \mathcal{U}_s\\
                  \delta \in \mathbb{R}}} u_s^{\top}u_s + \delta^2 \\
\label{clf_robust_p}
&\text{\st{} }s^{\top}(2\alpha_{\ell} (\mathcal{H}+M)+\dot{M}+MA+A^{\top}M+2\alpha_{\gamma} I)s \nonumber \\
&~~~~+2s^{\top}(\mathcal{B}+MB)u_s \leq \delta
\end{align}
where $M$ is given by (\ref{op_convex}) and the dependence on $x = [q^{\top},\dot{q}^{\top}]^{\top}$ is omitted for notational simplicity. Then we have
\begin{align}
\label{sboundclf}
E[\|s\|^2] \leq \frac{V(x(0),s(0))e^{-2\overline{\alpha}t}+\frac{C_{\ell}+\delta}{2 \overline{\alpha}}}{\inf_{t \geq 0}\lambda_{\min}(\mathcal{H}(q)+M(x))}.
\end{align}
where $V(x,s) = s^{\top}(\mathcal{H}(q)+M(x))s$ ($\sigma = 1$ is used in (\ref{lag_lyap})). Also, we can use $\delta = 0$ when $\mathcal{U}_s = \mathbb{R}^m$.
\end{proposition}
\begin{IEEEproof}
As in the proof of Theorem~\ref{thm_lag_sto} with $\sigma = 1$, we have
\begin{align}
\mathscr{L}V \leq& -2s^{\top}Ks+ s^{\top}(\dot{M}+MA+A^{\top}M+2\alpha_{\gamma} I)s \nonumber \\
&+2s^{\top}(\mathcal{B}+MB)u_s+C_{\ell}. 
\end{align}
This inequality with the condition (\ref{clf_robust_p}) gives $\mathscr{L}V\leq2\overline{\alpha}V+C_{\ell}+\delta$, which yields (\ref{sboundinf}) by Theorem~\ref{sic}. The last part of this proposition follows from the fact that $u_s = -K_s s$ is a feasible solution of (\ref{clfqp_incon}) when $\mathcal{U}_s = \mathbb{R}^m$ and $\delta = 0$ for $M$ given by solving (\ref{op_convex}).
\end{IEEEproof}
\begin{remark}
The decision variable $\delta$ is introduced to avoid infeasibility due to the input constraint $u_s \in \mathcal{U}_s$. Also, for $\mathcal{U}_s = \mathbb{R}^m$, (\ref{clfqp_incon}) reduces to a quadratic program and has a computationally-efficient analytical solution~\cite{citeulike:163662}.
\end{remark}
\subsection{Finite-Dimensional Formulation of (\ref{convex_opt_gen}) and (\ref{op_convex})}
In order to solve (\ref{convex_opt_gen}) and (\ref{op_convex}), we need $\dot{\tilde{W}}$, $\overline{m}_{x}$, $\overline{m}_{x^2}$, and $\phi$ at each time instant. Assuming that an initial value of $\tilde{W}$ is given, $\dot{\tilde{W}}$ can be computed by backward difference approximation, $\dot{\tilde{W}}(t_k) \simeq (\tilde{W}(x(t_k))-\tilde{W}(x(t_{k-1})))/dt$, where $\tilde{W}(x(t_k))$ is a decision variable of the current convex optimization problem and $\tilde{W}(x(t_{k-1}))$ is a given constant as a result of the convex optimization at the previous time step $t_{k-1}$. We can perform similar operations for computing $\overline{m}_{x}$, $\overline{m}_{x^2}$, and $\phi$ at each time instant.

For practical applications, it is also possible to neglect them or assign approximate values to each variable~\cite{observer}, although the resultant parameters could be sub-optimal in these cases.
\subsection{Computationally-Efficient CV-STEM Algorithms}
Since solving (\ref{convex_opt_gen}) or (\ref{op_convex}) at every time step can be computational intractable for some systems, we propose several ways to update the contraction metric less frequently.
\subsubsection{Relaxed CV-STEM Algorithm}
\label{relaxation1}
This method updates the control parameters only when one of the constraints in (\ref{convex_opt_gen}) or (\ref{op_convex}) is violated, or when the objective value at the current iteration is larger than that at the previous iteration. Since this will not change the stability proof, the controller still guarantees exponential boundedness of the mean squared tracking error of system trajectories. This approach will be demonstrated in Sec.~\ref{simulation} along with the discussion on how to select the sampling period $\Delta t$ of the CV-STEM control.
\subsubsection{Approximate CV-STEM Algorithm}
We could approximate the sampled CV-STEM solutions offline assuming the form of a contraction metric in a given hypothesis function space. One candidate is the polynomial basis function space, which leads to the sum-of-squares programming-based search algorithm~\cite{1425952,AYLWARD20082163,5443730}. However, its application is limited by the facts that it is developed for systems with a polynomial vector field and that the problem size grows exponentially with the number of variables and basis functions~\cite{sos_dissertation}. We also have several machine-learning based techniques for numerically modeling the CV-STEM sampled optimal contraction metrics~\cite{ncm,nscm}.
\subsection{Coefficients of SDC Parameterizations}
There are two variations of (\ref{convex_opt_gen}) with the relaxed constraints (\ref{riccati_lmi_con_rho}), (\ref{alpha_cond_convex_2_rho}), and (\ref{last_constraint_con}) in Proposition~\ref{relaxed_convex}, when selecting $\varrho$ of SDC matrices.
We can either set them to some given values \textit{a priori} to preserve the controllability, or pre-compute a constant solution $M$ offline using constant parameterizations of $A$~\cite{observer}.
\section{Numerical Simulation}
\label{simulation}
The performance of the CV-STEM is evaluated in the following two problems, where convex optimization problems are solved using \textit{cvx} toolbox in Matlab~\cite{cvx, gb08}. Since running an optimization algorithm at every time step is unrealistic in practice, the relaxed CV-STEM in Sec.~\ref{variations} is used in this section along with the discussion on the sampling period $\Delta t$ introduced in Theorem~\ref{disc_cont_thm}. The computation of $d{\tilde{W}}/dt$ is performed by backward difference approximation. A Matlab implementation of the CV-STEM algorithm is available at {\color{caltechgreen}\underline{\url{https://github.com/astrohiro/cvstem}}}.
\subsection{Spacecraft Attitude Control}
We first consider the spacecraft attitude dynamics given in~\cite{alt_controller,simulation} with stochastic disturbances.
\subsubsection{Simulation Setup}
The spacecraft state (modified Rodrigues parameters) is initialized as ${q}(0) = [0.9,-0.9,0.7]^{\top}$, ${\dot{q}}(0) = [0.6,0.7,-0.5]^{\top}$, and $G_u(x,t)$ in (\ref{original_dynamics}) is given as $G_u(x,t) = 0.2\times[0,0,0,1,1,1]^{\top}$. We initialize $\tilde{W}$ by solving the CV-STEM without the $d{\tilde{W}}/dt$ term.
The desired trajectories are defined as
${q}_{1d} = 0.3\sin(2\pi(0.1)t)$, ${q}_{2d} = 0.2\sin(2\pi(0.2)t+\pi/6)$, and ${q}_{3d} = 0$ and the CV-STEM is applied with $\alpha = 10^{-3}$ and $R = I$. The input constraint in Proposition~\ref{tot_input_con_prop} is used with $u_{\max} = 700$. The same simulation is performed for PID, $\mathcal{H}_{\infty}$~\cite{256331}, and a nonlinear controller with an exponential stability guarantee~\cite{alt_controller}, where the PID gains are selected as $K_P = 1300I$, $K_I = 300I$ and $K_D = 1300I$. We use $K_r = 100I$ and $\Lambda = I$ for the controller in~\cite{alt_controller}. The sampling period $\Delta t = 0.1$ is used for the CV-STEM and $\mathcal{H}_{\infty}$ control.
\subsubsection{Simulation Results}
\begin{figure}
    \centering
    \includegraphics[width=72mm]{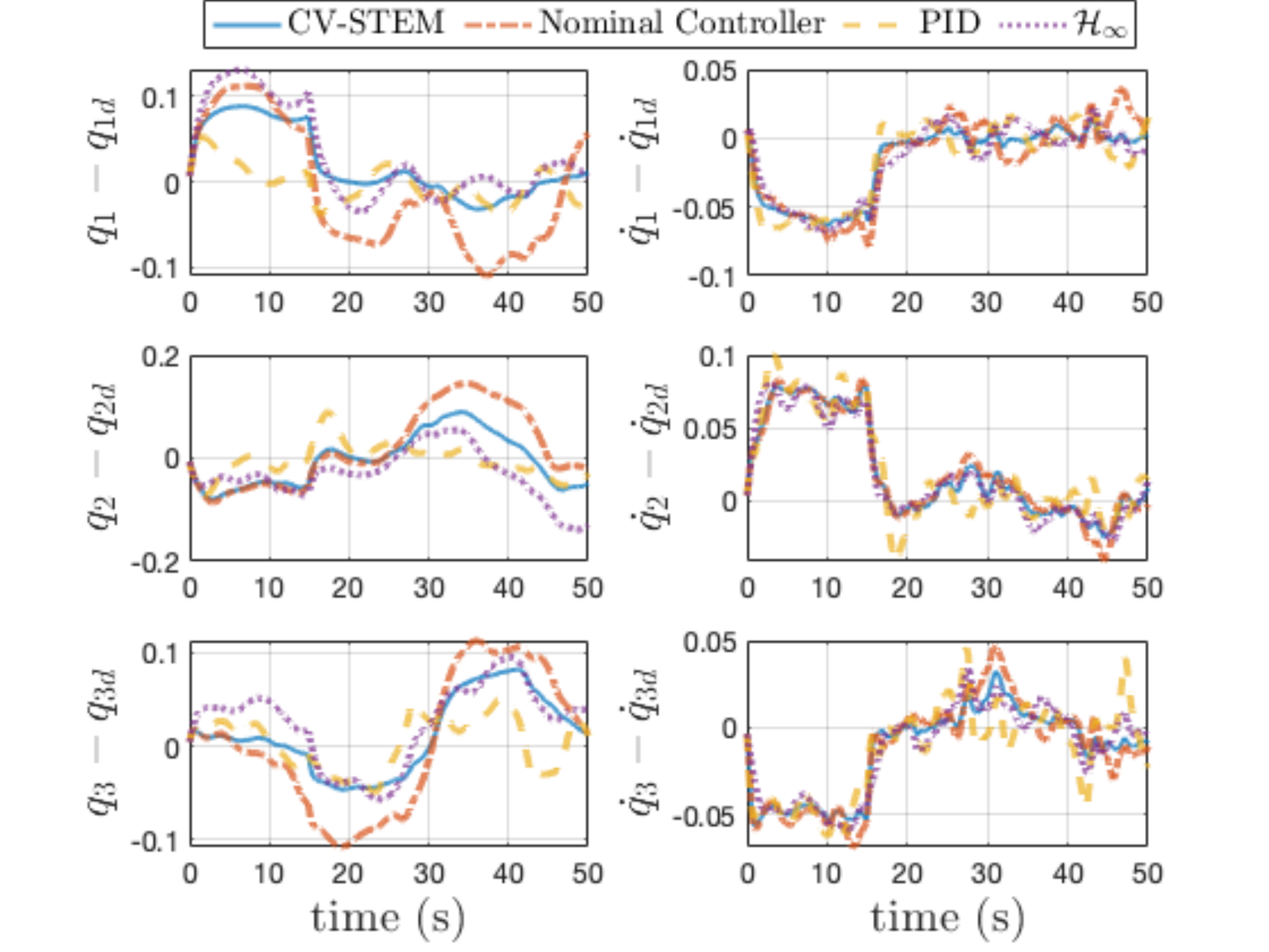}
    \caption{Tracking errors of Modified Rodrigues parameters}
    \label{mrp}
\end{figure}

\begin{figure}
    \centering
    \includegraphics[width=72mm]{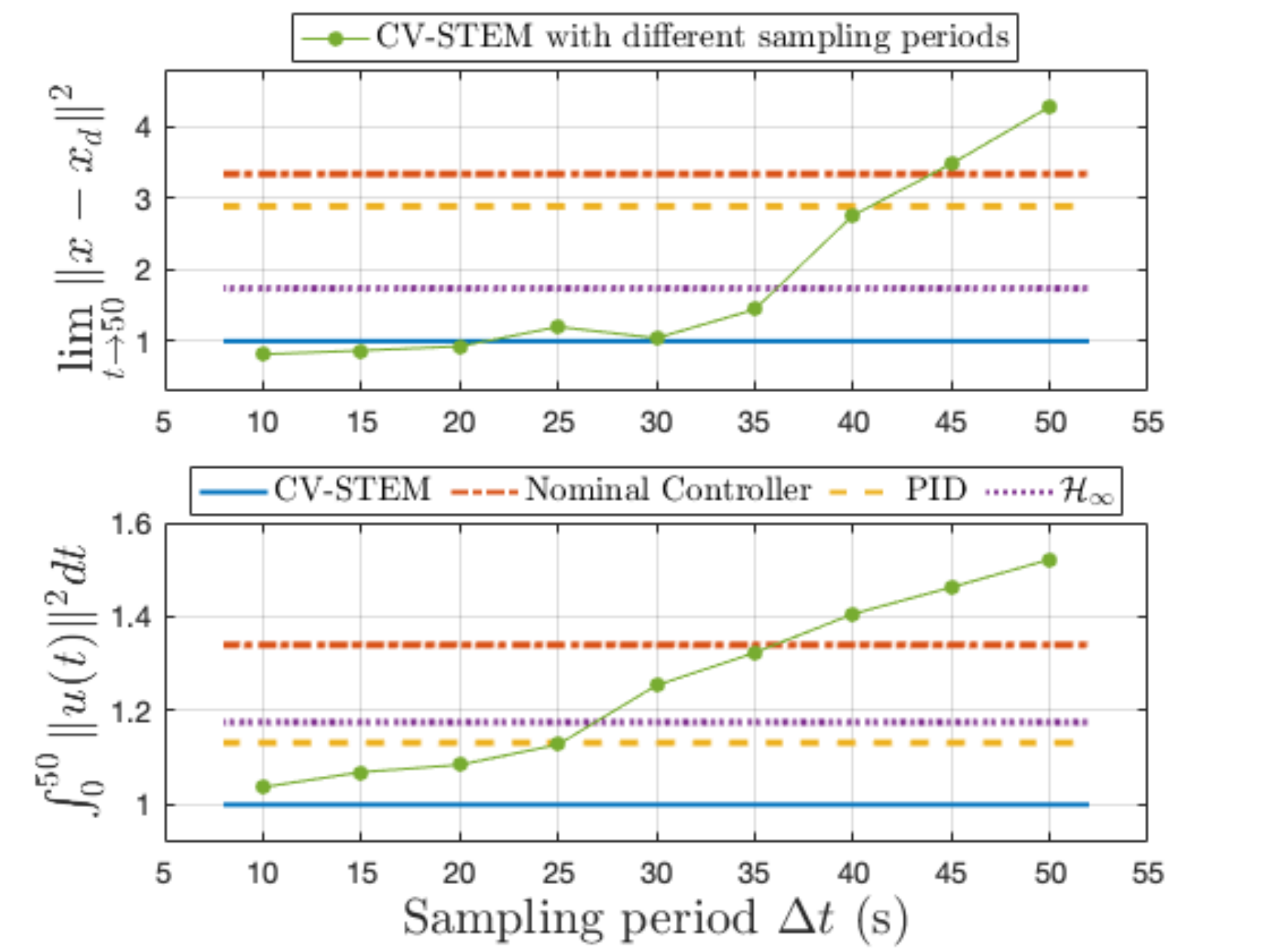}
    \caption{Steady-state tracking errors and control effort for spacecraft attitude control: Values in the figure are computed by the average over $60$ simulations and normalized by one at the CV-STEM performances. The steady-state error is computed by the average over the values of last 150 steps at each simulation to account for the stochasticity in the system.}
    \label{mse_and_control}
\end{figure}

Figure~\ref{mrp} shows tracking errors of each state for the CV-STEM, the controller in~\cite{alt_controller}, PID, and $\mathcal{H}_{\infty}$ control, smoothed by the 150-point moving average filter. Figure~\ref{mse_and_control} shows the normalized steady-state tracking error $\lim_{t\to50}\|x(t)-x_d(t)\|^2$ and control effort $\int_{0}^{50}u(t)dt$ of each controller averaged over $60$ simulations, where $x = [q^{\top},\dot{q}^{\top}]^{\top}$. It also includes those of the CV-STEM control with different sampling periods $\Delta t$ to see the impact of discrete-time implementation of the proposed algorithm. It should be noted $\lim_{t\to\infty}\|x(t)-x_d(t)\|^2$ is what we attempt to minimize. It is computed by the average over the values of last 150 steps at each simulation to account for the stochasticity in the system. Table~\ref{sim_result_1} summarizes the steady-state tracking error and control effort for each controller depicted as horizontal lines in Fig.~\ref{mse_and_control}.

It is shown that the proposed CV-STEM achieves a smaller steady-state tracking error than the controller in~\cite{alt_controller}, PID, and $\mathcal{H}_{\infty}$ control with a smaller amount of control effort as shown in Figs.~\ref{mrp}--\ref{mse_and_control} and Table~\ref{sim_result_1}. Also, the error of the CV-STEM with its sampling period $\Delta t \leq 35$ (s) remains smaller than the other three even with smaller control effort for $\Delta t \leq 25$ (s) as shown in Fig.~\ref{mse_and_control}. This fact implies that the CV-STEM control framework could be used in real-time with an onboard computer that solves the optimization within the period $\Delta t \leq 25, 35$ (s) whilst maintaining its superior performance. For example, solving the convex optimization takes less than $1.0$s with a Macbook Pro laptop (2.2 GHz Intel Core i7, 16 GB 1600 MHz DDR3 RAM).
\begin{table}
\caption{Control Performances for Spacecraft Attitude Control Computed as explained in Fig.~\ref{mse_and_control} \label{sim_result_1}}
\begin{center}
\begin{tabular}{|c|c|c|c|c|}
\hline
  & CV-STEM & Controller~\cite{alt_controller} & PID & $\mathcal{H}_{\infty}$ \\
\hline
\hline
Steady-state error & 1 & 3.3395 & 2.8849 & 1.7384 \\
Control effort & 1 & 1.3403 & 1.1319 & 1.1755 \\
\hline
\end{tabular}
\end{center}
\end{table}
\subsection{Multi-Agent System}
Next, we consider tracking and synchronization control of multiple formation flying spacecraft (5 agents) orbiting the earth. The detailed equation of motion and definition of symbols used in this simulation can be found in~\cite{CHUNG20131148}.
\subsubsection{Simulation Setup}
The desired trajectory of the leader agent is given as
$x_d(t) = 2.0\sin{(\omega t+\phi_{{e}_0})}$, $y_d(t) = 2.0\cos{(\omega t+\phi_{{e}_0})}$, and $z_d(t) = 0$.
See~\cite{CHUNG20131148} for how to construct synchronized desired orbits of the follower agents. We use $\Gamma(x,t) = [1,\cdots,1]^{\top} \in \mathbb{R}^{np\times 1}$ for the diffusion term defined in (\ref{sto_lag_system}), where $n=3$ ($3$ dimensional space) and $p=5$ ($5$ agents).
The tracking gain $K_1$ and the synchronization gain $K_2$ in~\cite{5068839} are selected as $K_1 = 5I$ and $K_2 = 2I$ with $\alpha = 10^{-3}$ and $R = I$ for the CV-STEM control. The spacecraft positions are initialized as uniformly distributed random variables over a cube with
side length $0.4$ ($-0.2 \leq x_j,y_j,z_j \leq 0.2$), velocities are as $[\dot{x}_j,\dot{y}_j,\dot{z}_j]^{\top} = [0,0,0]^{\top}$, and $\tilde{W}$ is as $\tilde{W}(0) = I$, for all agents $j$. The gain for the composite states in~\cite{5068839} is selected as $\Lambda_j = I,~\forall j$. Similar to the first simulation, the input constraint in Proposition~\ref{tot_input_con_prop} is used with $u_{\max} = 1.0$. 
For comparison, the nominal nonlinear controller in~\cite{5068839}, PID, and $\mathcal{H}_{\infty}$ control are also applied to this problem with $K_P = 7I$, $K_I = 0I$, and $K_D = 11I$. The sampling period $\Delta t = 0.5$ is used for the CV-STEM and $\mathcal{H}_{\infty}$.
\subsubsection{Simulation Results}
\begin{figure}
    \centering
    \includegraphics[width=72mm]{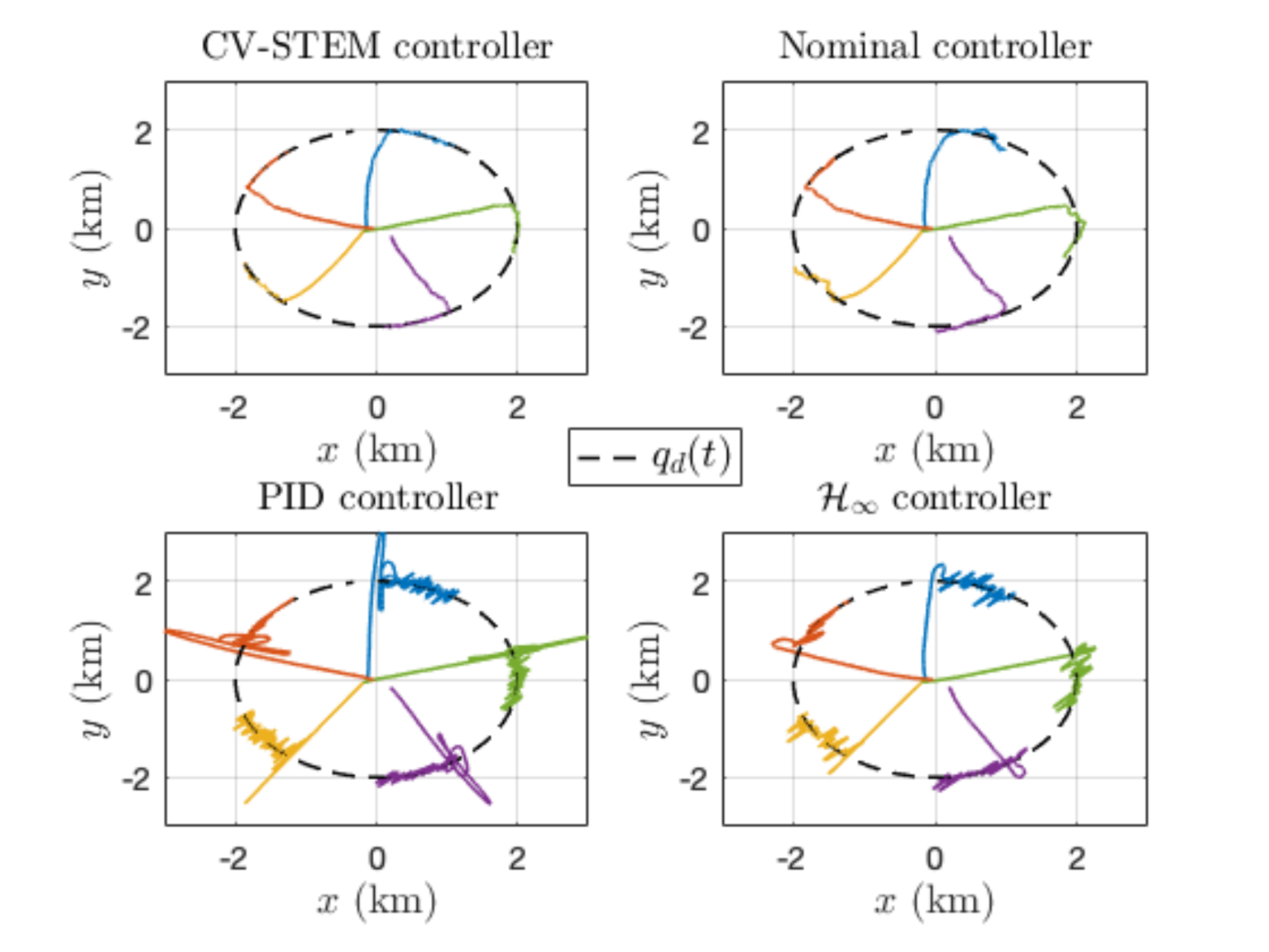}
    \caption{Controlled and desired spacecraft trajectories in the LVLH frame}
    \label{scmrp}
\end{figure}

\begin{figure}
    \centering
    \includegraphics[width=72mm]{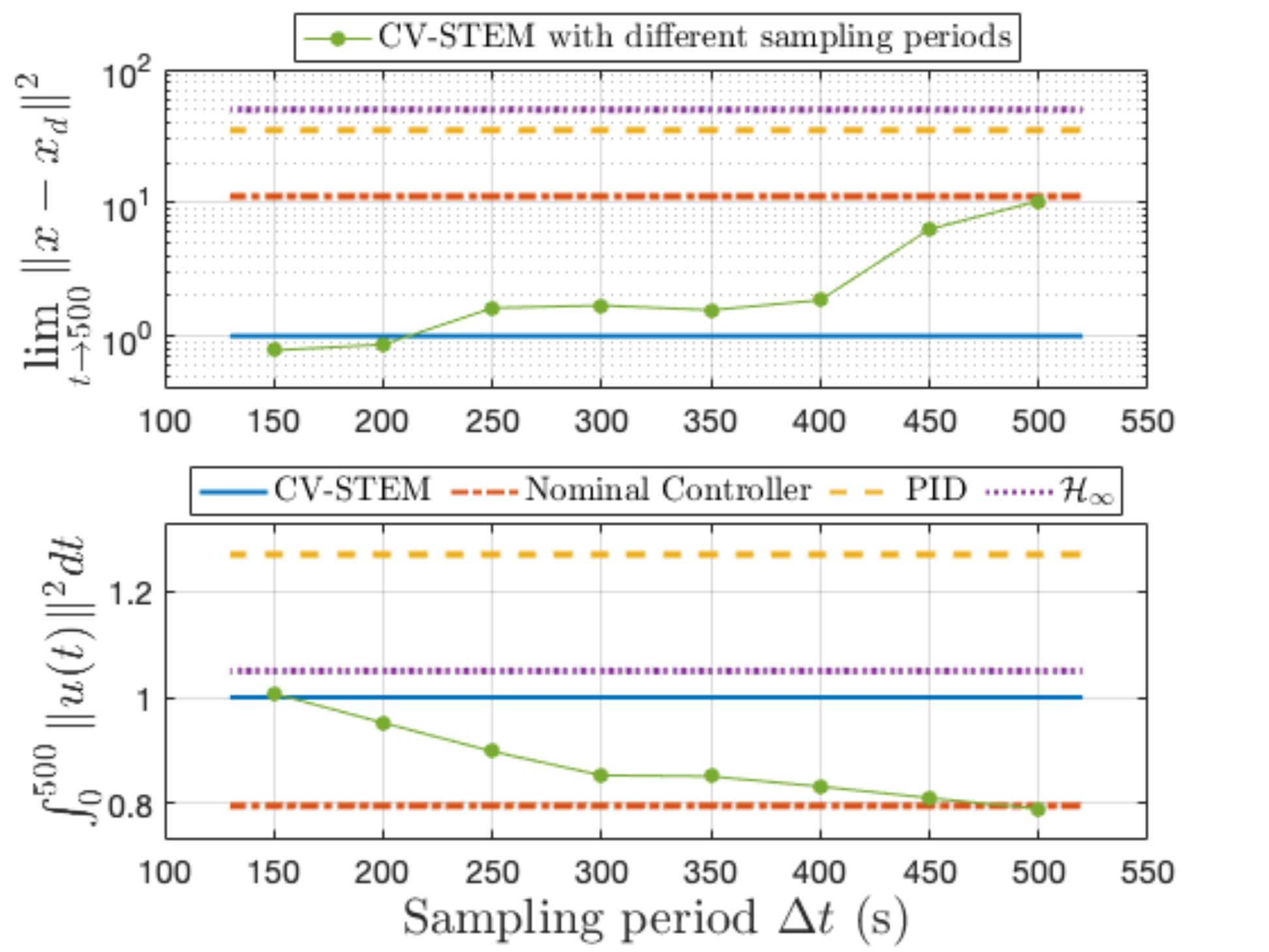}
    \caption{Steady-state tracking error and control effort for spacecraft tracking and synchronization control: Values in this figure are computed as explained in Fig.~\ref{mse_and_control}.}
    \label{multi2}
\end{figure}

Figure~\ref{scmrp} shows a comparison between the controlled and desired trajectories in the LVLH frame for the CV-STEM, the controller in~\cite{5068839,CHUNG20131148}, PID, and $\mathcal{H}_{\infty}$. Figure~\ref{multi2} shows the normalized steady-state tracking error and control effort of each controller and the CV-STEM with different sampling periods $\Delta t$, averaged over $60$ simulations. Again, the steady-state errors are computed by the average over the values of last 150 steps at each simulation. Table~\ref{sim_result_2} summarizes the control performances depicted as horizontal lines in Fig.~\ref{multi2}.

Figures~\ref{scmrp} and~\ref{multi2} indicate that the CV-STEM control performs better than the controller in~\cite{5068839,CHUNG20131148}, PID, and $\mathcal{H}_{\infty}$ control in terms of the steady-state tracking error. Due to the formulation $u = u_n+u_s$, its control effort is $1.25$ times larger than that of the nonlinear controller~\cite{5068839,CHUNG20131148} in this case as shown in Table~\ref{sim_result_2}. Furthermore, the error of the CV-STEM stays smaller than the others for the sampling period $\Delta t \leq 450$ (s) with control effort smaller than those of PID and $\mathcal{H}_{\infty}$ control. In particular, it is less than $1.7$ times as large as that of the nominal CV-STEM with $\Delta t=0.5$ (s) for $\Delta t \leq 350$ (s). This is a promising outcome for the real-time implementation of the CV-STEM control, as the aforementioned Macbook Pro laptop (2.2 GHz Intel Core i7, 16 GB 1600 MHz DDR3 RAM) solves the optimization within $1.5$s.
\begin{table}
\caption{Control Performances for Spacecraft Tracking and Synchronization Control Computed as Explained in Fig.~\ref{mse_and_control} \label{sim_result_2}}
\begin{center}
\begin{tabular}{|c|c|c|c|c|}
\hline
  & CV-STEM & Controller~\cite{5068839} & PID & $\mathcal{H}_{\infty}$ \\
\hline
\hline
Steady-state error & 1 & 11.176 & 34.997 & 49.903 \\
Control effort & 1 & 0.7946 & 1.2701 & 1.0496 \\
\hline
\end{tabular}
\end{center}
\end{table}
\section{Conclusion}
\label{conclusion}
In this paper, we present CV-STEM, a new numerical framework to construct an optimal contraction metric for feedback control of It\^{o} stochastic nonlinear systems and stochastic Lagrangian systems, expressed in SDC extended linear structure. It computes the metric by solving a convex optimization problem, which is proven to be equivalent to its non-convex counterpart of greedily minimizing an upper bound of the steady-state mean squared tracking error of the system trajectories. It is shown by stochastic incremental contraction analysis that the mean squared error is exponentially bounded for all time and for any initial condition, and that the CV-STEM control is robust against stochastic and deterministic disturbances. We also propose discrete-time stochastic contraction analysis with a state- and time-dependent metric to validate the sampling-based implementation of the algorithm. In numerical simulations, the CV-STEM control outperforms PID, $\mathcal{H}_{\infty}$, and nonlinear controllers developed for spacecraft attitude control and synchronization problems in terms of the steady-state tracking error, with the large enough sampling period which enables its real-time implementation.
\appendices
\section{Proof of Lemma~\ref{lemma_det_robustness_general}}
\label{proof_det_robustness}
\begin{IEEEproof}
Let us omit the arguments $x$ and $t$ for notational simplicity. Differentiating $V_M = e^{\top}Me$ with $e = x-x_d$ under the condition (\ref{riccati}) yields
\begin{align}
\dot{V}_M \leq e^{\top}(-\gamma M^2-MBR^{-1}B^{\top}M)e+2e^{\top}M(\Delta_d+Bd) \nonumber
\end{align}
where $\Delta_d = \Delta A x_d+\Delta B u_d$. Adding and subtracting $\|\mu_1\|=\|Sd\|^2$ where $R = S^2$ and completing the square, we have
\begin{align}
\dot{V}_M \leq& -\|y\|^2+\|\mu_1\|^2-\|\mu_1-S^{-1}B^{\top}Me\|^2+2e^{\top}M\Delta_d \nonumber.
\end{align}
where $y = (\sqrt{{\gamma}/{2}})M(x,t)e$. Using $\mu_2 = (\sqrt{{2}/{\gamma}})\Delta_d$,
\begin{align}
\dot{V}_M \leq& -\|y\|^2+\|\mu_1\|^2-\frac{1}{2}\gamma\left\|Me-\frac{2\Delta_d}{\gamma}\right\|^2+\frac{2}{\gamma}\|\Delta_d\|^2 \nonumber \\
\leq& -\|y\|^2+\|\mu_1\|^2+\|\mu_2\|^2.
\end{align}
By the comparison lemma~\cite[pp. 211]{Khalil:1173048}, this reduces to
\begin{align}
\|y_{\tau}\|_{\mathcal{L}_2} \leq \|(\mu_1)_{\tau}\|_{\mathcal{L}_2}+\|(\mu_2)_{\tau}\|_{\mathcal{L}_2}+\sqrt{V_M(x(0))}
\end{align}
which completes the proof.
\end{IEEEproof}
\section{Computation of $V_2$ and $\overline{V}_2$ in Theorem~\ref{thmcontracting}}
\label{computation_V2}
Using (\ref{l}), $V_2$ in Theorem~\ref{thmcontracting} can be computed as follows:
\begin{align}
&V_2 = \int_0^1\sum_{i,j}\frac{1}{2}\frac{\partial y}{\partial \mu}^{\top}M_{x_ix_j}\frac{\partial y}{\partial \mu}(G_uG_u^{\top})_{ij} \\
& +2(M_{i})_{x_j}\frac{\partial y}{\partial \mu}\left(G_u\frac{\partial G}{\partial \mu}^{\top}\right)_{ij}+m_{ij}\left(\frac{\partial G}{\partial \mu}\frac{\partial G}{\partial \mu}^{\top}\right)_{ij}d\mu \nonumber
\end{align}
where $M_i$ is the $i$th row of $M$ and the subscripts $x_i$ denote partial derivatives. Following the proof of Lemma~2 in~\cite{observer},
\begin{align}
\label{V2}
V_2 &\leq \overline{m}g_u^2+\int_0^12\overline{m}_{x}g_u^2\left\|\frac{\partial y}{\partial \mu}\right\|+\frac{1}{2}\overline{m}_{x^2}g_u^2\left\|\frac{\partial y}{\partial \mu}\right\|^2d\mu  \nonumber \\
&\leq 2\alpha_g\int_0^1\left\|\frac{\partial y}{\partial \mu}\right\|^2d\mu+\underline{m}C = \overline{V}_2
\end{align}
where $2\alpha_g = g_u^2\left(\overline{m}_{x}\varepsilon+\overline{m}_{x^2}/2\right)$ and $C = (\overline{m}/\underline{m})g_u^2+(\overline{m}_{x}g_u^2)/(\varepsilon\underline{m})$. The first inequality in (\ref{V2}) is due to $\trace(AB)\leq\|A\|\trace(B)$ for $A,B\succeq 0$, and the second inequality follows from the relation $2a'b' \leq \varepsilon^{-1}a'^2+\varepsilon b'^2$ for any scalars $a'$, $b'$, and $\varepsilon > 0$.
Thus, $V_2$ is upper bounded by $\overline{V}_2$ as desired.
\section*{Acknowledgment}
This work was in part funded by the Jet Propulsion Laboratory, California Institute of Technology and the Raytheon Company.
\bibliographystyle{IEEEtran}
\bibliography{ms}
\begin{IEEEbiography}[{\includegraphics[width=1in,height=1.25in,clip,keepaspectratio]{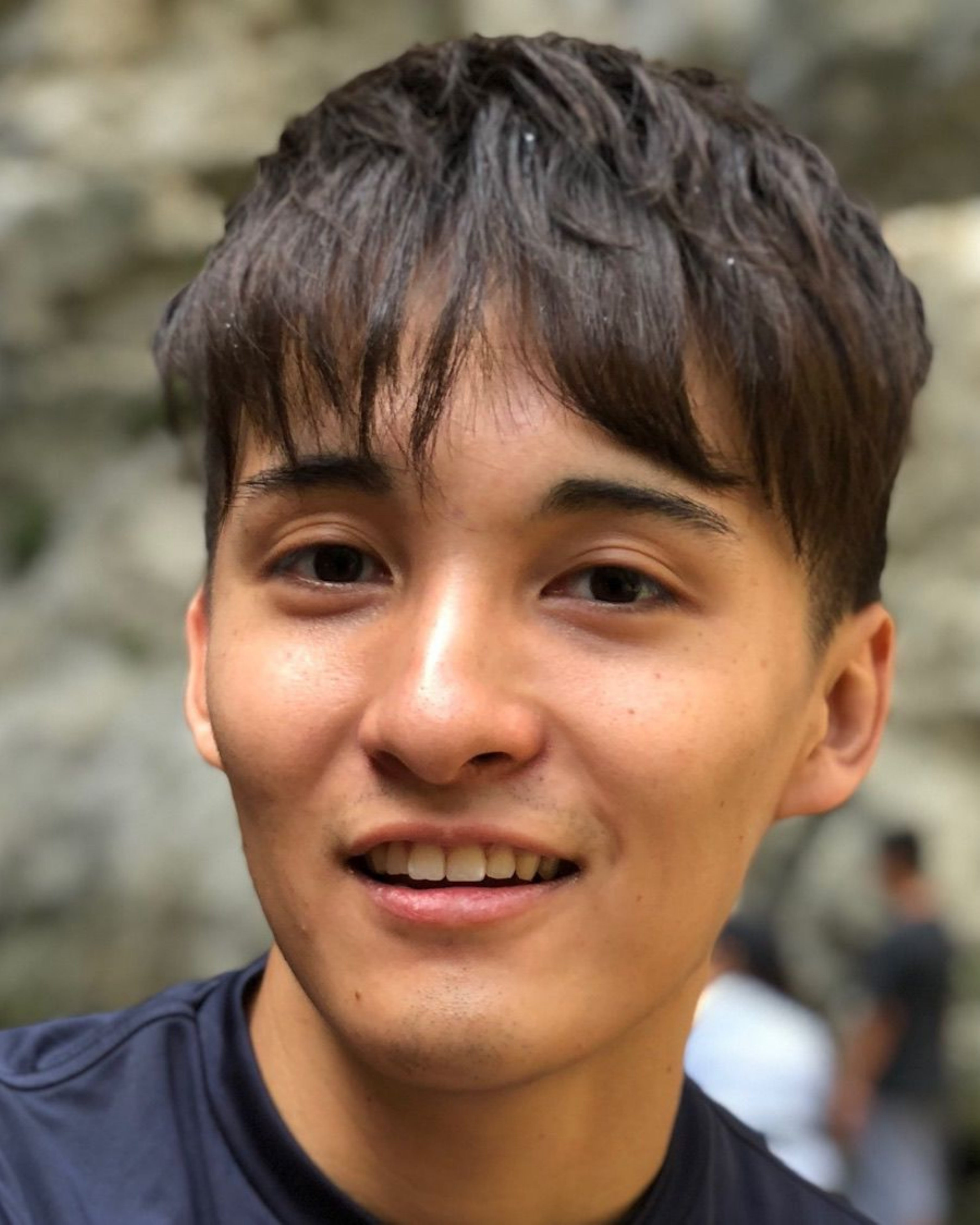}}]{Hiroyasu Tsukamoto} (M'19) received the B.S. degree in aerospace engineering from Kyoto University, Kyoto, Japan, in 2017 and the M.S. degree in space engineering from California Institute of Technology (Caltech), Pasadena, CA, USA, in 2018. He is currently pursuing the Ph.D. degree in space engineering at Caltech.
His research interests include systems and control theory, aerospace and robotic autonomy, and autonomous guidance, navigation, and control of general nonlinear systems with learning-based robustness, optimality, and stability guarantees. Mr. Tsukamoto is a recipient of the Caltech Vought Fellowship and the Funai Overseas Scholarship for graduate studies.
\end{IEEEbiography}
\begin{IEEEbiography}[{\includegraphics[width=1in,height=1.25in,clip,keepaspectratio]{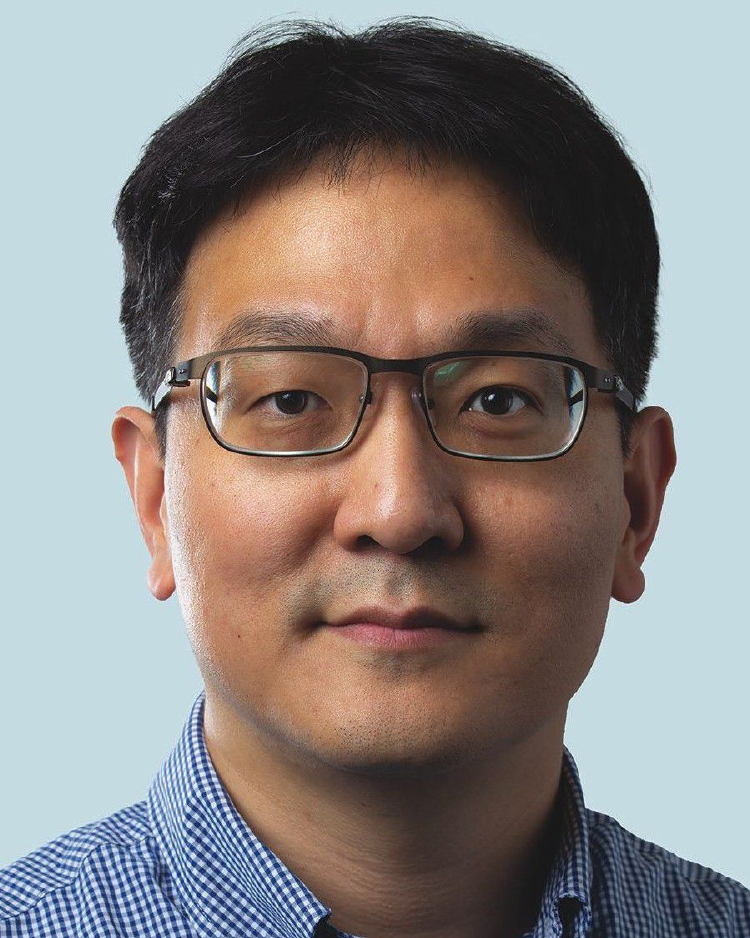}}]{Soon-Jo Chung} (M'06--SM'12) received the B.S. degree (\textit{summa cum laude}) in aerospace engineering from the Korea Advanced Institute of Science and Technology, Daejeon, South Korea, in 1998, and the S.M. degree in aeronautics and astronautics and the Sc.D. degree in estimation and control from Massachusetts Institute of Technology, Cambridge, MA, USA, in 2002 and 2007, respectively.

He is currently Bren Professor of Aerospace and a Jet Propulsion Laboratory Research Scientist in the California Institute of Technology, Pasadena, CA, USA. He was with the faculty of the University of Illinois at Urbana-Champaign (UIUC) during 2009–2016. His research interests include spacecraft and aerial swarms and autonomous aerospace systems, and in particular, on the theory and application of complex nonlinear dynamics, control, estimation, guidance, and navigation of autonomous space and air vehicles.

Dr. Chung was the recipient of the UIUC Engineering Deans Award for Excellence in Research, the Beckman Faculty Fellowship of the UIUC Center for Advanced Study, the U.S. Air Force Office of Scientific Research Young Investigator Award, the National Science Foundation Faculty Early Career Development Award, and three Best Conference Paper Awards from the IEEE and the American Institute of Aeronautics and Astronautics. He is an Associate Editor of IEEE Transactions on Robotics, IEEE Transactions on Automatic Control, and AIAA Journal of Guidance, Control, and Dynamics.
\end{IEEEbiography}
\end{document}